\documentclass[prb,onecolumn,showpacs,preprintnumbers,amsmath,assume]{revtex4}
\usepackage{graphicx}% Include figure files
\usepackage{dcolumn}% Align table columns on decimal point
\usepackage{bm}% bold math
\begin{document}

\title{Generalized Dix equation and analytic treatment \\ of normal-moveout velocity for anisotropic media}

\author{Vladimir Grechka$^{1,2}$, Ilya Tsvankin$^{1}$, and Jack K. Cohen$^{1}$}
\affiliation{$^{1}$Center for Wave Phenomena, Colorado School of Mines, Golden, CO  80401-1887} 
\affiliation{$^{2}$currently at Marathon Oil Company}
\date{\today}

\begin{abstract}

Despite the complexity of wave propagation in anisotropic media, reflection 
moveout on conventional common-midpoint (CMP) spreads is usually well 
described by the normal-moveout (NMO) velocity defined in the zero-spread 
limit. In their recent work, Grechka and Tsvankin showed that the azimuthal 
dependence of NMO velocity generally has an {\it elliptical}$\,$ shape and is 
determined by the spatial derivatives of the slowness vector evaluated at the 
CMP location. This formalism is used here to develop exact solutions for 
normal-moveout velocity in anisotropic media of arbitrary symmetry. 

For the model of a single homogeneous layer above a dipping reflector, we 
obtain an {\it explicit}$\,$ NMO expression valid for all pure modes and any 
orientation of the CMP line with respect to the reflector strike. The 
influence of anisotropy on normal-moveout velocity is absorbed by the 
slowness components of the zero-offset ray (along with the derivatives of the 
vertical slowness with respect to the horizontal slownesses) -- quantities 
that can be found in a straightforward way from the Christoffel equation. 
If the medium above a dipping reflector is horizontally stratified, the effective NMO velocity is determined through a Dix-type average of the matrices responsible for the ``interval'' NMO {\it ellipses}$\,$ in the individual layers. This generalized 
Dix equation provides an analytic basis for moveout inversion in vertically 
inhomogeneous, arbitrary anisotropic media. For models with a throughgoing 
vertical symmetry plane (i.e., if the dip plane of the reflector coincides 
with a symmetry plane of the overburden), the semi-axes of the NMO ellipse 
are found by the more conventional rms averaging of the interval NMO 
{\it velocities}$\,$ in the dip and strike directions.

Modeling of normal moveout in the most general heterogeneous anisotropic media 
requires {\it dynamic} ray tracing of only one (zero-offset) ray. 
Remarkably, the expressions for geometrical spreading along the zero-offset 
ray contain all the components necessary to build the NMO ellipse. This method 
is orders of magnitude faster than multi-azimuth, multi-offset ray tracing 
and, therefore, can be efficiently used in traveltime inversion and in devising fast dip-moveout (DMO) processing algorithms for anisotropic media. This algorithm becomes especially efficient if the model consists of homogeneous layers or blocks separated by smooth 
interfaces. 

The high accuracy of our NMO expressions is illustrated by comparison with ray-traced reflection traveltimes in piecewise-homogeneous, azimuthally anisotropic models. We also apply the generalized Dix equation to field data collected over a fractured reservoir and show that $P$-wave moveout can be used to find the depth-dependent fracture orientation and evaluate the magnitude of azimuthal anisotropy. 

\end{abstract}
 % 81.05.Xj for anisotropy, 91.30.-f for seismology
\maketitle

%%%%%%%%%%%%%%%%%%%%%%%%%%%%%%%%%%%%%%%%%%%%%%%%%%%%%%%%%%%%%%%%%%%%%%%%%%%%%%%

\section{Introduction}

Reflection moveout in inhomogeneous anisotropic media is usually calculated 
by multi-offset and multi-azimuth ray tracing \cite{GajewskiPsencik1987}. 
%(e.g., Gajewski and P\v{s}en\v{c}\'{\i}k 1987). 
While the existing anisotropic ray-tracing codes are 
sufficiently fast for forward modeling, their application in moveout inversion 
requires repeated generation of azimuthally-dependent traveltimes around many 
common-midpoint (CMP) locations, which makes the inversion procedure extremely 
time-consuming. Moveout modeling, however, can be simplified by taking advantage of the 
limited range of offsets in conventional acquisition design. For common 
spreadlength-to-depth ratios close to unity, CMP traveltimes in media with 
moderate structural complexity are well described by the normal-moveout (NMO) 
velocity defined in the zero-spread limit\cite{ TsvankinThomsen1994, GrechkaTsvankin1998}. Even if the data exhibit nonhyperbolic moveout, 
NMO velocity is still responsible for the most stable, small-offset portion 
of the moveout curve. 

Existing methods for computing normal-moveout velocity in inhomogeneous media are designed for isotropic models\cite{Shah1973, HubralKrey1980}. 
Angular velocity variations make both analytic and 
computational aspects of NMO-velocity modeling much more complicated. Here, 
we present a treatment of NMO velocity in inhomogeneous anisotropic 
media that provides an analytic basis for moveout inversion, leads to a dramatic increase in the efficiency of traveltime modeling methods, and helps to develop insight into the influence of the anisotropic parameters on reflection traveltimes.

Explicit expressions for normal-moveout velocity are well known for the 
relatively simple transversely isotropic model with a vertical symmetry axis\cite{Thomsen1986} 
(VTI media). Recently, Tsvankin\cite{Tsvankin1995}
presented an exact NMO equation for dipping reflectors valid in vertical 
symmetry planes of any homogeneous anisotropic medium. Alkhalifah and 
Tsvankin\cite{AlkhalifahTsvankin1995} extended this result by developing a Dix-type equation for 
vertically {\it inhomogeneous} anisotropic media above a dipping reflector. 
They also showed that the NMO-velocity function in VTI media depends on just 
two parameters -- the zero-dip NMO velocity $V_{\rm nmo} (0)$ and the 
``anellipticity'' coefficient $\eta$. Still, their formalism is limited to 
2-D wave propagation in the dip plane of the reflector, which should also 
coincide with a symmetry plane of the overburden. 

This work is based on  a general 3-D treatment of normal moveout developed by 
Grechka and Tsvankin\cite{GrechkaTsvankin1998}, who proved that the azimuthal dependence of NMO velocity for pure (non-converted) modes has an {\it elliptical}$\,$ shape in the 
horizontal plane, even if the medium is arbitrary anisotropic and 
inhomogeneous. This conclusion breaks down only for subsurface models in which 
common-midpoint reflection traveltime cannot be described by a series 
expansion or does not increase with offset. The orientation of the NMO ellipse 
and the values of its semi-axes can be expressed through the spatial 
derivatives of the slowness vector, which are determined by both the direction 
of the reflector normal and the medium properties above the reflector. 

Grechka and Tsvankin\cite{GrechkaTsvankin1998} also presented explicit representations of the NMO velocity for a horizontal orthorhombic 
layer and dipping reflectors beneath VTI media. A detailed analysis of the NMO ellipse for transversely isotropic media with a horizontal symmetry axis (HTI media) is given in Tsvankin\cite{Tsvankin1997}, who also discusses the inversion of conventional-spread reflection moveout for the parameters of HTI media. Sayers\cite{SayersSEG1995} obtained the elliptical dependence of NMO velocity for the model of a homogeneous anisotropic layer with a horizontal symmetry plane using an approximation for long-spread moveout based on group-velocity expansion in spherical harmonics. 
 
Here, we apply the formalism of Grechka and Tsvankin\cite{GrechkaTsvankin1998} to more 
complicated anisotropic models. We start by deriving an explicit expression 
for azimuthally-dependent NMO velocity from a dipping reflector overlaid by 
a homogeneous layer of arbitrary symmetry. Then we obtain a generalized Dix 
equation for NMO velocity in a model composed of a stack of horizontal 
homogeneous, arbitrary-anisotropic layers above a dipping reflector. While 
this equation has a form similar to the conventional Dix formula, it is based 
on the averaging of the {\it matrices} that define interval NMO {\it ellipses}. 
For the most general inhomogeneous media, we develop an efficient methodology to compute the normal-moveout 
velocity using the dynamic ray-tracing 
of only \emph{one} (zero-offset) ray. We show that the derivatives needed to 
find the geometrical spreading\cite{CervenyMolotkovPsencik1977, KendallThomson1989} 
%(e.g., \v{C}erveny, Molotkov and P\v{s}en\v{c}\'{\i}k 1977; Kendall and Thomson 1989) 
provide sufficient 
information to build the NMO ellipse and, therefore, model reflection moveout 
without tracing a large family of rays. Finally, we compare the hyperbolic moveout equation parameterized by the exact NMO velocity with ray-traced reflection traveltimes and present a field-data application of the generalized Dix differentiation. 
  
%=============================================================================%
\section{Equation of the NMO ellipse}

\begin{figure}
\includegraphics[width = 6.5 cm]{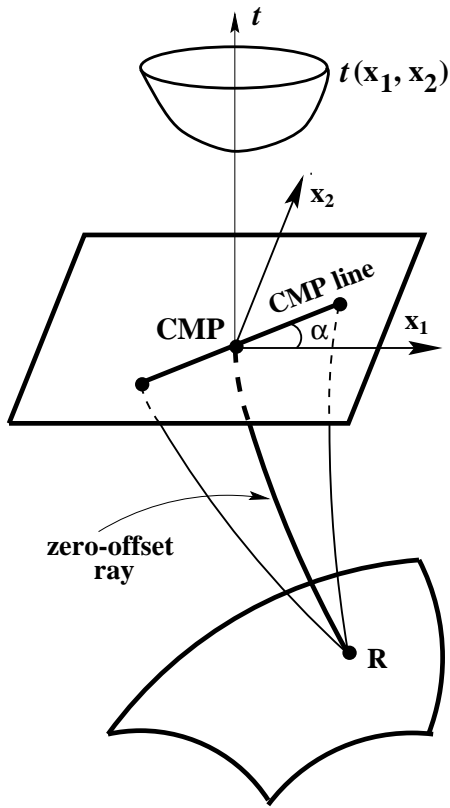} \vspace{-3mm}
%\centerline{\epsfxsize=6.5cm \epsffile{fig1.eps}}
\caption{Normal-moveout velocity is calculated on CMP lines with different azimuths and a fixed midpoint location. It is not necessary to account for reflection-point dispersal in the derivation of NMO velocity.}
\label{fig1}
\end{figure}

Suppose the traveltimes of a certain reflected wave (reflection moveout) 
have been recorded on a number of common-midpoint (CMP) gathers with different 
azimuthal orientation but the same midpoint location (Figure~\ref{fig1}). If the medium is anisotropic and inhomogeneous, the dependence of large-offset reflection 
traveltimes on the azimuth $\alpha$ of the CMP line may become rather 
complicated. For conventional spreadlengths close to the distance between the 
CMP and the reflector, however, moveout in CMP geometry is usually 
well-approximated by a hyperbolic equation,
\begin{equation}
   t^2(\alpha)\approx t^2_0 \, + \, \frac{x^2}{V^2_{\rm nmo} \, (\alpha)} \, .
   \label{iteq:v1}
\end{equation}
Here $t_0$ is the zero-offset reflection time, $x$ is the source-receiver 
offset, and $V_{\rm nmo} \, (\alpha)$ is the normal-moveout velocity defined 
as 
\begin{equation}
   V_{\rm nmo}^2 \, (\alpha) = 
   \lim_{x \rightarrow 0} \, {d[x^2] \over d[t^2(\alpha)]} \, .
   \label{iteq:v7}
\end{equation}

The analysis 
below is based on the general result of Grechka and Tsvankin\cite{GrechkaTsvankin1998}, who 
showed that NMO velocity is described by the following simple quadratic form:
\begin{equation}
      V_{\rm nmo}^{-2} (\alpha) = 
                 W_{11} \cos^2 \alpha +
            2 \, W_{12} \sin \alpha \cos \alpha +
                 W_{22} \sin^2 \alpha \, ,
  \label{eq009}
\end{equation}
where ${\bf W}$ is a symmetric matrix defined as 
\begin{equation}
      W_{i j} 
          = \tau_0 \, { {\partial^2 \tau} \over 
                        {\partial x_i \partial x_j} } 
                        \Biggl|_{\bf x_{\rm CMP}}
          = \tau_0 \, { {\partial p_i} \over {\partial x_j} } 
                        \Biggl|_{\bf x_{\rm CMP}} , \quad (i,j = 1,2) \, .
  \label{eq0091}
\end{equation}
Here, $\tau (x_1, \, x_2)$ is the {\it one-way} traveltime from the 
zero-offset reflection point to the location ${\bf x} \, \{x_1, \, x_2\}$ at 
the surface, $\tau_0$ is the one-way zero-offset traveltime, $p_i$ are the 
components of the slowness vector corresponding to the ray emerging at the 
point {\bf x}, and ${\bf x_{\rm CMP}}$ corresponds to the CMP location; the origin of the coordinate system in the derivations below coincides with the zero-offset reflection point. The one-way traveltimes appear in equation~(\ref{eq0091}) 
because reflection-point dispersal has no influence on the NMO velocity of 
pure modes, and (for the small source-receiver offsets appropriate for
estimation of $V_{\rm nmo}$) rays can be assumed to propagate through the 
reflection point of the zero-offset ray\cite{HubralKrey1980, Tsvankin1995}.

It is convenient to use the eigenvectors of the matrix {\bf W} as auxiliary 
horizontal axes and rotate the NMO equation~(\ref{eq009}) by the angle 
$\beta$ (see Appendix~A), 
\begin{equation}
      \beta = \tan^{-1} \left[ \frac{W_{22} - W_{11}  + 
                  \sqrt{ (W_{22} - W_{11})^2 + 4 W_{12}^2 }} 
              {2 W_{12}} \right] .
   \label{eq0621}
\end{equation}

This rotation transforms equation~(\ref{eq009}) into 
\begin{equation}
      V_{\rm nmo}^{-2} (\alpha) = \lambda_1 \cos^2 (\alpha - \beta) +
                                  \lambda_2 \sin^2 (\alpha - \beta) \, ,
  \label{b001}
\end{equation}
where $\lambda_{1,2}$ are the eigenvalues of the matrix {\bf W}.
Grechka and Tsvankin\cite{GrechkaTsvankin1998} conclude that for positive $\lambda_1$ and 
$\lambda_2$ the NMO velocity~(\ref{eq009}) represents an {\it ellipse} in 
the horizontal plane. A negative eigenvalue implies a negative 
$V^2_{\rm nmo}$ in certain azimuthal directions and, consequently, a decrease 
in the CMP traveltime with offset. Although such reverse moveout can exist 
in some cases (e.g., for turning waves, as described by Hale et al.\cite{Haleetal1992}), 
typically both $\lambda_1$ and $\lambda_2$ are positive, and the azimuthal 
dependence of NMO velocity is indeed elliptical. Note that this conclusion 
is valid for arbitrary inhomogeneous anisotropic media provided the 
traveltime field is sufficiently smooth to be adequately approximated by a 
Taylor series expansion. 

%=============================================================================%
\section{Homogeneous arbitrary anisotropic layer}

\subsection{General case}

To obtain normal-moveout velocity for any given model from 
equations~(\ref{eq009}) and~(\ref{eq0091}), we need to evaluate the spatial 
derivatives of the slowness vector at the CMP location. As demonstrated in 
Appendix~B, for the model of a single homogeneous layer this can be done by 
representing the horizontal ray displacement through group velocity and using 
the relation between the group-velocity and slowness vectors. As a result, 
we find the following explicit expressions for the matrix {\bf W} and 
azimuthally-dependent NMO velocity [equations~(\ref{eq032}) and~(\ref{eq633})]: 
\begin{equation}
  {\bf W} = { {p^{}_1 q^{}_{,1} + p^{}_2 q^{}_{,2} - q} \over 
              {q^{}_{,11} q^{}_{,22} - q_{,12}^2} } \, 
          \begin{pmatrix}
			   q^{}_{,22} & - q^{}_{,12} \cr
			  - q^{}_{,12} &   q^{}_{,11} \cr 
          \end{pmatrix} \! ,
  \label{eq932}
\end{equation}

\begin{eqnarray}
  V_{\rm nmo}^{-2} (\alpha) & \equiv & V_{\rm nmo}^{-2} (\alpha, p^{}_1, p^{}_2)
  \nonumber \\ & = & {{p^{}_1 q^{}_{,1} + p^{}_2 q^{}_{,2} -q} \over     {q^{}_{,11} q^{}_{,22} - q_{,12}^2} } \, \left[ q^{}_{,22} \cos^2 \alpha - 2 q^{}_{,12} \sin \alpha
    \cos \alpha + q^{}_{,11} \sin^2 \alpha \right] ,
  \label{eq033}
\end{eqnarray}
where $q \equiv q(p^{}_1, p^{}_2) \equiv p^{}_3$ denotes the vertical slowness 
component, $q^{}_{,i} \equiv {\partial q} / {\partial p^{}_i}$, and 
$q^{}_{,ij} \equiv {\partial^2 q} / {\partial p^{}_i} {\partial p^{}_j}$; the 
horizontal components of the slowness vector ($p^{}_1$ and $p^{}_2$) and the derivatives in 
equation~(\ref{eq033}) are evaluated for the zero-offset ray.

Equation~(\ref{eq033}) is valid for pure modes reflected from horizontal or 
dipping interfaces in media with arbitrary symmetry and any strength of the 
anisotropy. The normal-moveout velocity is fully determined by the azimuth
$\alpha$ of the CMP line and the slowness vector of the 
zero-offset ray. The slowness components $p_1$, $p_2$ and $q$ can be found by solving the Christoffel equation for the slowness (phase) direction normal to the reflector. (The slowness vector of the zero-offset ray is normal to the reflecting interface at the reflection point.) Since this equation is cubic with respect to the squared phase velocity, it yields an explicit expression for the slowness vector. 

The derivatives of the vertical slowness $q$ can be found directly from the Christoffel equation as well. As discussed in more detail below in the section on ray tracing, the slowness components satisfy the equation $F(q, p^{}_1, p^{}_2)=0$, where $F$ is (in general) a sixth-order polynomial 
with respect to $q$. For common anisotropic models with a horizontal symmetry plane (e.g., the medium can be transversely isotropic, orthorhombic or even monoclinic), $F$ becomes a {\it cubic} polynomial with respect to $q^2$.
Hence, the derivatives $q_{,i}$ and $q_{,ij}$ can be obtained as
\[
      q_{,i} = - { {F_{p_i}} \over {F_q} }
\]
and 
\begin{equation}
      q_{,ij} = - { {F_{p_i p_j} + F_{p_i q} q_{,j} + F_{p_j q} q_{,i} + 
                    F_{qq} q_{,i} q_{,j} } \over {F_q} } \, ,
  \label{eq034}
\end{equation}
where $F_{p_i} \equiv {\partial F} / {\partial p_i}$,
$F_q \equiv {\partial F} / {\partial q}$,
$F_{p_i p_j} \equiv {\partial^2 F} / {\partial p_i} {\partial p_j}$,
$F_{p_i q} \equiv {\partial^2 F} / {\partial p_i} {\partial q}$, and
$F_{qq} \equiv {\partial^2 F} / {\partial q^2}$. 
Therefore, all terms in equation~(\ref{eq033}) can be obtained {\it explicitly}$\,$ from the Christoffel equation. 

Equation~(\ref{eq033}) can also be used to develop weak-anisotropy approximations for NMO velocity by linearizing $q$ and its derivatives in dimensionless anisotropic parameters or in perturbations in the stiffness coefficients. These analytic approximations provide valuable insight into the influence of the anisotropic parameters on normal moveout\cite{Tsvankin1995, Cohen1998}. There is hardly any need, however, to substitute weak-anisotropy approximations for the exact equations in numerical modeling.

\begin{figure}
\includegraphics[width = 3.6 in]{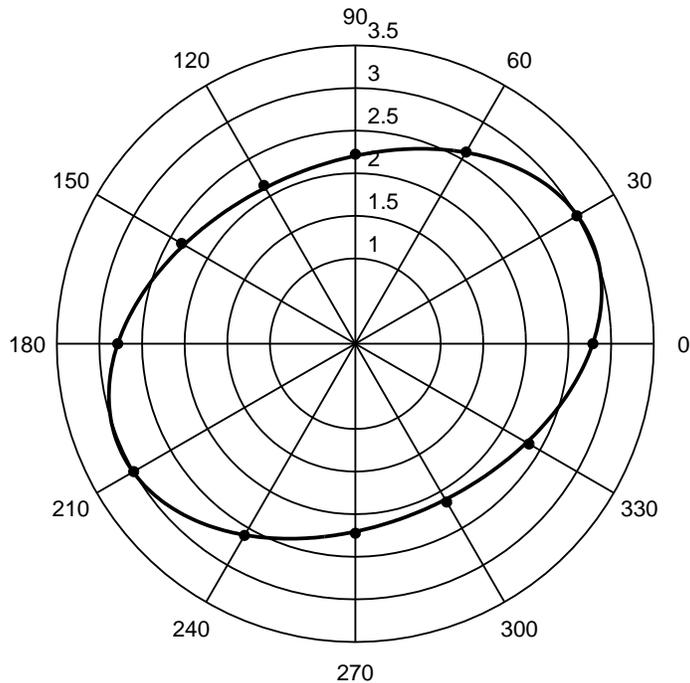} \vspace{-5mm}
%\centerline{\epsfxsize=3.6in \epsffile{Fig04.eps}}
\caption{Comparison of the $P$-wave NMO velocity from equation 
         (\protect\ref{eq033}) (solid line) and the moveout (stacking) 
         velocity (dots) obtained by least-squares fitting of a hyperbola to 
         the exact traveltimes computed for spreadlength equal to the distance between the CMP and the reflector. The model contains a homogeneous orthorhombic layer (with the vertical symmetry planes at azimuths 0$^\circ$ and 90$^\circ$)
         above a plane dipping reflector; the dip and azimuth of the 
         reflector are equal to 30$^\circ$ (azimuthal angles are 
         shown along the perimeter of the plot). The relevant medium parameters 
         [in Tsvankin's\cite{Tsvankin1997} notation] are $V_{P0}=2.0$ km/s, 
         $\epsilon^{\rm (1)}=0.110$, $\delta^{\rm (1)}=-0.035$, 
         $\epsilon^{\rm (2)}=0.225$, $\delta^{\rm (2)}=0.100$, 
         $\delta^{\rm (3)}=0$. The vertical symmetry plane at zero 
         azimuth has the properties of the VTI model of Dog Creek shale, 
         while the second vertical symmetry plane is equivalent to Taylor 
         sandstone; both models are described in Thomsen\cite{Thomsen1986}.}
\label{fig04}
\end{figure}

Thus, equation~(\ref{eq033}) gives a simple and numerically efficient 
recipe to obtain azimuthally-dependent reflection moveout in an arbitrary 
anisotropic layer. The example in Figure~\ref{fig04}, generated for an orthorhombic layer above a dipping reflector, illustrates the high accuracy of the hyperbolic moveout equation parameterized by the analytic NMO 
velocity~(\ref{eq033}) in describing conventional-spread reflection 
traveltimes. Despite the 
presence of anisotropy-induced nonhyperbolic moveout, the $P$-wave NMO 
velocity is close to the moveout (stacking) velocity calculated from the 
exact traveltimes on six CMP lines with different orientation. The maximum 
difference between $V_{\rm nmo}$ (solid line) and the finite-spread moveout 
velocity (dots) is just 1.4\%, which is even less than the corresponding 
value (2.7\%) for the same model, but with a {\it horizontal} reflector\cite{GrechkaTsvankin1998}. Therefore, the magnitude of nonhyperbolic 
moveout for this model decreases with reflector dip; the same observation 
was made by Tsvankin\cite{Tsvankin1995} for vertical transverse isotropy. Note that 
although the azimuth of the dip plane of the reflector is equal to 
30$^\circ$, the largest axis of the $V_{\rm nmo}(\alpha)$ ellipse has an 
azimuth of 24.3$^\circ$ due to the influence of the azimuthal anisotropy 
above the reflector. 

%------------------------------------------------------------------------------

\subsection{Special cases}

\subsubsection{Model with a vertical symmetry plane}
Next, let us consider a special case -- a model in which the dip plane of the 
reflector coincides with a vertical symmetry plane of the layer. The medium can be, for instance, transversely isotropic with the symmetry axis confined to the dip plane or orthorhombic. The mirror 
symmetry with respect to the dip plane implies that one of the axes of the 
NMO ellipse points in the dip direction. Below, we provide a formal proof 
of this fact, as well as concise expressions for the azimuthally dependent 
NMO velocity in this model.   

It is convenient to align the $x_1$-axis with the azimuth of the dip plane, 
while the $x_2$-axis will point in the strike direction. Evidently, the 
zero-offset ray should lie in the vertical symmetry plane $x_2 = 0$, and 
its slowness component $p^{}_2$ goes to zero. As another consequence of the 
mirror symmetry with respect to the dip plane, 
$\partial p^{}_2 / \partial x_1 = 0$ (i.e, rays corresponding to $x_2=0$ stay in the dip plane and cannot have a non-zero $p^{}_2$), so the cross-term 
$q_{,12}$ in equation~(\ref{eq033}) vanishes, and the NMO velocity simplifies 
to
\begin{equation}
      V_{\rm nmo}^{-2} (\alpha, p^{}_1) =
        {{p^{}_1 q^{}_{,1} - q}  \over {q^{}_{,11} q^{}_{,22}} } \,
            \left[ q^{}_{,22} \cos^2 \alpha 
                 + q^{}_{,11} \sin^2 \alpha \right]  \! .
  \label{eq0345}
\end{equation}
Equation~(\ref{eq0345}) describes an ellipse with the semi-axes in the dip ($\alpha=0$) and strike ($\alpha=\pi/2$) directions:
\begin{equation}
      V_{\rm nmo}^2 (\alpha = 0, \, p^{}_1) = { {q^{}_{,11}} \over {p^{}_1 q^{}_{,1} - q} } \, ,
  \label{eq0346}
\end{equation}
\begin{equation}
      V_{\rm nmo}^2 (\alpha = {\pi \over 2}, \, p^{}_1 ) = 
        { {q^{}_{,22}} \over {p^{}_1 q^{}_{,1} - q} } \, .
  \label{eq0347}
\end{equation}

The dip-line NMO velocity~(\ref{eq0346}) was originally obtained via the 
in-plane phase velocity $V$ by Tsvankin\cite{Tsvankin1995}:
\begin{equation}
   V_{\rm nmo}(0, \phi) = \frac{V(\phi)}{\cos\phi} \, 
                          \dfrac{\sqrt{1 + \left. \dfrac{1}{V(\phi)}
                                                  \dfrac{d^2V}{d\theta^2} \right|_{\,\theta=\phi} }}
                               {1 - \left. \dfrac{\tan\phi}{V(\phi)} \,
                                     \dfrac{dV}{d\theta}\right|_{\,\theta=\phi} }  \, , 
   \label{eq999}
\end{equation}
where $\theta$ is the phase angle with vertical in the dip plane, and $\phi$ is the reflector dip. In the form~(\ref{eq0346}) $V_{\rm nmo}^2 (0, p^{}_1)$ was first given by Cohen\cite{Cohen1998}. Equation~(\ref{eq0347}) provides a similar representation for the NMO velocity in the strike direction. 

Equations~(\ref{eq0346}) and~(\ref{eq0347}) are always valid for transversely 
isotropic media with a vertical symmetry axis because of the mirror symmetry 
with respect to {\it any} vertical plane in this model. The vertical slowness 
in VTI media can be represented as 
$q (p^{}_1,p^{}_2) \equiv q \! \left( \! \sqrt{p_1^2 + p_2^2}\right)$ and, 
for $p^{}_2 = 0$, $q^{}_{,22} = q^{}_{,1}/p^{}_1$. Then equation~(\ref{eq0347}) 
for the strike-line NMO velocity reduces to the expression obtained 
previously\cite{GrechkaTsvankin1998},
\begin{equation}
      V_{\rm nmo}^2 (\alpha = {\pi \over 2}, p^{}_1) =
        { {q^{}_{,1}} \over {p^{}_1 (p^{}_1 q^{}_{,1} - q)} } \, .
  \label{eq0348}
\end{equation}
Grechka and Tsvankin\cite{GrechkaTsvankin1998} also gave an equivalent form of 
equation~(\ref{eq0348}) in terms of the phase-velocity function and the 
weak-anisotropy approximation for $V_{\rm nmo}^2 (\alpha = {\pi \over 2}, p^{}_1)$. Due to the axial symmetry of the VTI model, both the dip-line [equation~(\ref{eq0346})] and strike-line 
[equation~(\ref{eq0348})] NMO velocities depend on the derivatives of $q$ 
with respect to the single horizontal (in-plane) slowness component ($p^{}_1$). 
The cubic equation for $q^2 (p^{}_1)$ in VTI media is particularly easy to solve 
because it splits into a quadratic equation for $P-SV$ waves and a linear 
equation for the $SH$-wave.  

Finally, in isotropic media the vertical slowness can be directly expressed 
through the reflector dip $\phi$: 
\[
q = \sqrt{ V^{-2} - p_1^2 } = {\cos \phi \over V} \, ,
\] 
and equations~(\ref{eq999}) and (\ref{eq0348}) yield the well-known
relationships presented by Levin\cite{Levin1971}:
\begin{equation}
      V_{\rm nmo}(\alpha = 0) = { V \over {\cos \phi} } \, ,
  \label{eq0349}
\end{equation}
\begin{equation}
      V_{\rm nmo}(\alpha = {\pi \over 2}) = V \, . 
  \label{eq0440}
\end{equation}

\subsubsection{Horizontal reflector}
For a horizontal reflector ($p_1=p_2=0$), equation~(\ref{eq033}) reduces to
\begin{equation}
  \label{eq662}
 V_{\rm nmo}^{-2} (\alpha,0,0)   = - \, {  {q}  \over { q^{}_{,11} q^{}_{,22} - q_{,12}^2} } \, \left[ q^{}_{,22} \cos^2 \alpha - 2 q^{}_{,12} \sin \alpha
    \cos \alpha + q^{}_{,11} \sin^2 \alpha \right] ,  
\end{equation}
where $q$ and $q^{}_{,ij}$ should be evaluated at the vertical slowness direction. 

Further simplification can be achieved for a medium with a vertical symmetry plane. Aligning the $x_1$-axis with the symmetry-plane direction and substituting $p_1=0$ into equation~(\ref{eq0345}) [or $q_{,12}=0$ into equation~(\ref{eq662})] yields 

\begin{equation}
      V_{\rm nmo}^{-2} (\alpha) = - \, 
        { {q} \over  {q^{}_{,11} q^{}_{,22}}} \,
            \left[ q^{}_{,22} \cos^2 \alpha 
                 + q^{}_{,11} \sin^2 \alpha \right] .
  \label{eq0945}
\end{equation}
As shown by Grechka and Tsvankin\cite{GrechkaTsvankin1998}, for an orthorhombic layer (that has two mutually orthogonal symmetry planes) $P$-wave NMO velocity from equation~(\ref{eq0945}) becomes a simple function of the vertical $P$-wave velocity $V_{P0}$ and the anisotropic coefficients $\delta^{(1)}$ and $\delta^{(2)}$ defined by Tsvankin\cite{Tsvankin1997}:
   
\begin{equation}
   V^2_{\rm nmo}(\alpha) = V^2_{P0} \, \frac{(1 +2\delta^{(1)}) \, 
                                  (1 +2\delta^{(2)})} 
             {1 +2\delta^{(2)} \sin^2\alpha + 2\delta^{(1)} \cos^2\alpha} \, .
\label{iteq:v929}
\end{equation}
Note that the linearized $\delta$ coefficients introduced by Mensch and Rasolofosaon\cite{MenschRasolofosaon1997} and Gajewski and P\v{s}en\v{c}\'{\i}k\cite{GajewskiPsencik1996} within the framework of the weak-anisotropy approximation are not appropriate for the exact equation~(\ref{iteq:v929}). Normal-moveout velocities for vertical and horizontal transverse isotropy can be easily found as special cases of equation~(\ref{iteq:v929})\cite{GrechkaTsvankin1998}. Equation~(\ref{eq0945}) can also be used to derive similar expressions for the split shear waves in orthorhombic media. 

%=============================================================================%

\section{Horizontally-layered medium above a dipping reflector}

\subsection{Generalized Dix equation}

\begin{figure}
	\includegraphics[width = 6.5 cm]{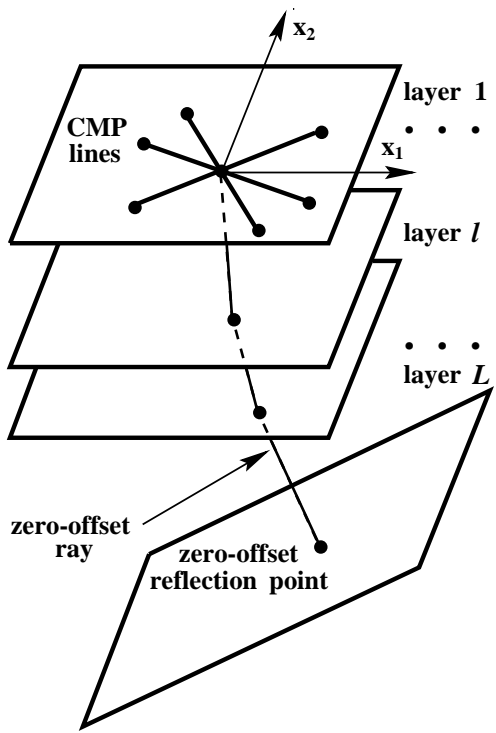} \vspace{-5mm}
%\centerline{\epsfxsize=7cm  \epsffile{laymodel.eps}}
\caption{A dipping reflector beneath a horizontally layered overburden. Normal-moveout velocity in this model can be obtained from the generalized Dix equation derived here.}
\label{laymodel}
\end{figure}

Here, we show that the NMO ellipse for vertically inhomogeneous arbitrary 
anisotropic media above a dipping  reflector (Figure~\ref{laymodel}) can be obtained by Dix-type averaging of the matrices {\bf W} responsible for the interval NMO ellipses. In our derivation we essentially follow the approach employed by Alkhalifah and Tsvankin\cite{AlkhalifahTsvankin1995} to obtain a ``2-D'' Dix-type NMO equation for rays confined to the incidence (vertical) plane that contains the CMP line. Their equation is valid only in the dip plane of the reflector, which should 
also coincide with a symmetry plane of the medium. In contrast, we make no 
assumptions about the mutual orientation of the CMP line and reflector strike,
and take full account of the out-of-plane phenomena associated with both 
model geometry and depth-varying anisotropy.

To construct the effective NMO ellipse, we need to obtain the matrix {\bf W} defined 
in equation~(\ref{eq0091}):
\begin{equation}
   W_{i j}(L) = \tau(L) {\partial p_i \over \partial x_j(L)} \, , 
               \quad (i,j = 1,2) \, ,
\end{equation}
where $\tau(L)$ is the total zero-offset traveltime and $x_i(L)$ is the 
horizontal ray displacement between the zero-offset reflection point located 
at the $L$-th (generally dipping) interface and the surface (Figure~\ref{laymodel}). Due to the 
continuity of the ray, both $\tau(L)$ and $x_i(L)$ are equal to the sum of 
the respective interval values: 
\begin{equation}
      \tau(L) = \sum_{\ell=1}^L \tau_\ell \, ,
  \label{eq0253}
\end{equation}
\begin{equation}
      x_i(L) = \sum_{\ell=1}^L x_{i,\ell} \, , \quad (i = 1,2).
  \label{eq0251}
\end{equation}
(Note that here and below in the section on layered media, the comma in the subscripts separates the layer index and {\it does not} denote differentiation.) 

It is convenient to introduce an auxiliary matrix 
\begin{equation}
      Y_{i j}(L) \equiv { {\partial x_i(L)} \over {\partial p_j} } 
                \, , \quad (i,j = 1,2)
  \label{eq0252a}
\end{equation}
with derivatives evaluated for the ray parameters $p^{}_1$ and $p^{}_2$ of the 
zero-offset ray. Then 
\begin{equation}
      {\bf W} \equiv {\bf W}(L) 
              = \tau(L) \, {\bf Y}^{-1}(L) \, .
  \label{eq0254}
\end{equation}
In a model composed of horizontally homogeneous layers 
above the reflector, the horizontal components $p^{}_1$ and $p_2$ of the slowness vector remain constant along any given ray between the reflection point and the surface. Therefore, substituting equation~(\ref{eq0251}) into 
equation~(\ref{eq0252a}), we find 
\begin{equation}
 \label{eq0252}
     Y_{i j}(L) \equiv { {\partial x_i(L)} \over {\partial p_j} }=
               \sum_{\ell=1}^L { {\partial x_{i,\ell}} \over {\partial p_j} }
               \equiv \sum_{\ell=1}^L Y_{ij,\ell} \,.
\end{equation}
Equation~(\ref{eq0252}) explains the reason for introducing the effective 
matrix ${\bf Y}(L)$: unlike the matrix {\bf W}, it can be decomposed into 
the sum of the matrices ${\bf Y}_\ell$ for the individual layers. Since all 
intermediate boundaries are horizontal, the ray displacements $x_{i,\ell}$ 
in any layer coincide with the values that should be used in computing the 
matrix {\bf W} and the interval NMO velocity for this particular layer. 
Hence, we can apply equation~(\ref{eq0254}) to layer $\ell$:
\begin{equation}
      {\bf W}_\ell = \tau_\ell \, {\bf Y}_\ell^{-1} \,  
  \label{eq0255}
\end{equation}
and, therefore, 
\begin{equation}
      {\bf Y}_\ell = \tau_\ell\, {\bf W}_\ell^{-1} \, .
  \label{eq0257}
\end{equation}
Substituting equations~(\ref{eq0257}) and~(\ref{eq0252})
into equation~(\ref{eq0254}) leads to the final result:
\begin{equation}
      {\bf W}^{-1}(L) = { 1 \over {\tau(L)} } \, 
             \sum_{\ell=1}^L \tau_\ell \, {\bf W}_\ell^{-1} \, .
  \label{dixavr}
\end{equation}
Interval matrices ${\bf W}_\ell$ in terms of the components of the slowness 
vector are given by equation~(\ref{eq932}), while the traveltimes 
$\tau_\ell$ should be obtained from the kinematic ray tracing (i.e., by 
computing group velocity) of the zero-offset ray. Note that, since the 
eigenvalues of the matrices ${\bf W}_\ell$ and ${\bf W}(L)$ usually are 
positive (under the assumptions discussed in Grechka and Tsvankin\cite{GrechkaTsvankin1998}), 
these matrices are nonsingular and, therefore, can be inverted. 

Equation~(\ref{dixavr}) performs Dix-type averaging of the interval 
matrices ${\bf W}_\ell$ to obtain the effective matrix ${\bf W}(L)$ and the effective normal-moveout velocity $V_{\rm nmo}(\alpha,L)$. 
It should be emphasized that the interval NMO velocities 
$V_{{\rm nmo},\ell}(\alpha)$ (or the interval matrices ${\bf W}_\ell$) in 
equation~(\ref{dixavr}) are computed for the horizontal components of the 
slowness vector of the zero-offset ray. (As follows from Snell's law, the 
slowness vector of the zero-offset ray is normal to the reflector at the 
reflection point.) This means that the interval matrices ${\bf W}_\ell$ in 
equation~(\ref{dixavr}) correspond to the generally {\it non-existent} 
reflectors that are orthogonal to the slowness vector of the zero-offset 
ray in each layer. 

Rewriting equation~(\ref{dixavr}) in the Dix 
differentiation form gives 
\begin{equation}
      {\bf W}_\ell^{-1} = 
             { { \tau(\ell) {\bf W}^{-1}(\ell) -
                 \tau(\ell-1) {\bf W}^{-1}(\ell-1) } \over
               { \tau(\ell) - \tau(\ell-1) } } \, .   
  \label{eq0260}
\end{equation}

Equations~(\ref{dixavr}) and~(\ref{eq0260}) generalize the Dix\cite{Dix1955}
formula for horizontally-layered arbitrary anisotropic media above a dipping
reflector. Formally, this extension looks relatively straightforward: the 
squared NMO velocities in the Dix formula are simply replaced by the inverse 
matrices ${\bf W}^{-1}$. Also, the generalized Dix differentiation is subject to the same limitation as its conventional counterpart: the thickness of the layer of interest (in vertical time) should not be too much smaller than the layer's depth.  

In contrast to the conventional Dix equation, 
however, the effective matrix ${\bf W}^{-1}(\ell-1)$ in equation~(\ref{eq0260}) {\it cannot}$\,$ be obtained 
from seismic data directly since the corresponding reflector usually does 
not exist in the subsurface. Therefore, layer-stripping by means of 
equation~(\ref{eq0260}) involves recalculating each interval matrix ${\bf W_\ell}$ from one value of the slowness vector (corresponding to a certain real 
reflector in a given layer) to another -- that of the zero-offset ray. 
This procedure was discussed for the 2-D case by Alkhalifah and Tsvankin\cite{AlkhalifahTsvankin1995} and further developed for $P$-waves in VTI media by Alkhalifah\cite{Alkhalifah1997}; the 
latter paper also contains a successful application of this algorithm to 
field data. 

Only in the simplest special case of a {\it horizontal}$\,$ reflector, does the 
slowness vector of the zero-offset ray not change its direction (stays 
vertical) all the way to the surface, and the interval matrices 
${\bf W}_{\ell}$ correspond to the NMO velocities from horizontal interfaces 
that can be measured from reflection data. Note that although such a model 
is horizontally-homogeneous, the zero-offset {\it ray} is not necessarily vertical
(if the medium does not have a horizontal symmetry plane), and the zero-offset reflection point may be shifted in the horizontal direction from the CMP location.

%------------------------------------------------------------------------------

\subsection{Model with a vertical symmetry plane}

Next, we consider the same special case as for the single-layer model -- a medium in which all layers have 
a common vertical symmetry plane that coincides with the dip plane of the 
reflector (e.g., the symmetry is VTI). For such a model the matrices ${\bf W}_\ell$ in the individual 
layers are diagonal (see the previous section), and
\begin{equation}
      W_{12,\ell} = 0 \, .
  \label{eq02611}
\end{equation}
Consequently, the off-diagonal elements of the matrix ${\bf W}(L)$ 
[equation~(\ref{dixavr})] vanish as well:
\begin{equation}
      W_{12}(L) = 0 \, .
  \label{eq02612}
\end{equation}
If the matrix {\bf W} is diagonal, its two components directly determine the 
semi-axes of the NMO ellipse [see equation~(\ref{eq009}) and Appendix~A]:
\begin{equation}
      W_{kk,\ell} = [ V_{{\rm nmo},\ell}^{(k)} ]^{-2}
  \label{eq0262}
\end{equation}
and 
\begin{equation}
      W_{kk}(L) = [ V_{\rm nmo}^{(k)}(L) ]^{-2} \, , \quad  (k=1,2) \, ,
  \label{eq0263}
\end{equation}
where [$V_{\rm nmo}^{(1)} \equiv V_{\rm nmo}(\alpha = 0)$] and 
[$V_{\rm nmo}^{(2)} \equiv V_{\rm nmo}(\alpha = {\pi / 2})$] are the NMO 
velocities measured in the dip and strike directions, respectively.

Substitution of equations~(\ref{eq02611}) -- (\ref{eq0263}) into 
equations~(\ref{dixavr}) and (\ref{eq0260}) yields more conventional Dix-type 
averaging and differentiation formulas for the dip- and strike-components of 
the normal-moveout velocity:
\begin{equation}
      [ V_{\rm nmo}^{(k)}(L) ]^2 = {1 \over {\tau(L)} }
      \sum_{\ell=1}^L \tau_\ell \, [ V_{{\rm nmo},\ell}^{(k)} ]^2
  \label{eq0264}
\end{equation}
and 
\begin{equation}
      [ V_{{\rm nmo},\ell}^{(k)} ]^2 = 
      { {\tau(\ell) [ V_{\rm nmo}^{(k)}(\ell) ]^2 -
         \tau(\ell-1) [ V_{\rm nmo}^{(k)}(\ell-1) ]^2} \over 
        { \tau(\ell) - \tau(\ell-1)} } \, , \quad (k=1,2) \, .
  \label{eq0265}
\end{equation}

Equations~(\ref{eq0264}) and~(\ref{eq0265}) for the {\it dip} component ($k=1$)
of the NMO velocity were derived by Alkhalifah and Tsvankin\cite{AlkhalifahTsvankin1995}
who considered 2-D wave propagation in the dip plane of the reflector. 
Our derivation shows that the same Dix-type equations can be applied to the 
{\em strike}-component ($k=2$) of the NMO velocity, which determines the second 
semi-axis of the NMO ellipse. Despite the close resemblance of 
expressions~(\ref{eq0264}) and (\ref{eq0265}) to the conventional Dix 
equation, the interval NMO velocities in equations~(\ref{eq0264}) 
and~(\ref{eq0265}), as in the more general Dix equation discussed above, 
correspond to the {\it non-existent}$\,$ reflectors normal to the slowness vector 
of the zero-offset ray in each layer.

%------------------------------------------------------------------------------

\subsection{Accuracy of the rms averaging of NMO velocities}

Although the generalized Dix equation~(\ref{dixavr}) operates with the 
{\em matrices}$\,$ ${\bf W}_\ell^{-1}$, we proved that Dix-type averaging can be applied to the dip- and strike-components of the {\em normal-moveout velocity}$\,$ in a model that has a common (throughgoing) vertical symmetry plane aligned with the dip plane of the reflector. It is also clear from the results of the previous section that the rms averaging of the interval NMO velocities is valid in any azimuthal direction, if all interval NMO ellipses degenerate into circles. Hence, the error of this more conventional averaging procedure depends on the elongation of the interval ellipses, a quantity controlled by both azimuthal anisotropy and reflector dip. In Appendix~C we show that this error increases rather slowly as the interval ellipses pull away from a circle because the rms averaging of the interval velocities [equation~(\ref{d002})] provides a {\it linear} approximation $\tilde V_{\rm nmo}$ to the exact NMO velocity, if both are expanded in the ``elongation'' coefficient. 

\begin{figure}
\includegraphics[width = 3.6 in]{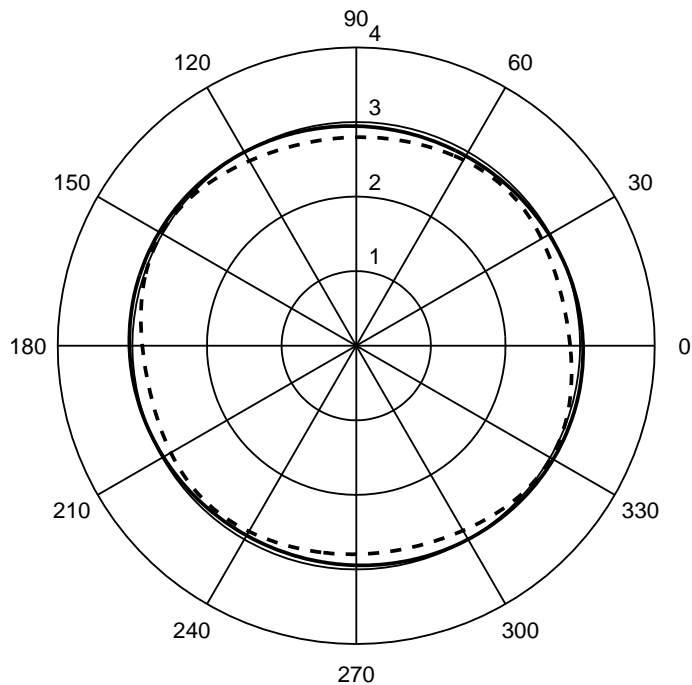} \vspace{-5mm}
%\centerline{\epsfxsize=3.6in \epsffile{Fig08.eps}}
\caption{Comparison of the exact $P$-wave NMO ellipse (solid line) and 
         an approximate NMO velocity obtained by the 
         Dix-type averaging [equation~(\protect\ref{d002}), dashed line]. 
         The model contains three horizontal orthorhombic layers with a   
         horizontal ($[x_1,x_2]$) symmetry plane. The azimuth of the 
         $[x_1,x_3]$ symmetry plane (also, the direction of one of the axes 
         of the interval NMO ellipse) in the first (subsurface) layer is 
         $\beta_1 = 0^\circ$, in the second layer -- $\beta_2 = 45^\circ$, 
         and in the third layer -- $\beta_3 = 60^\circ$. The vertical $P$-wave 
         velocities are $V_{P0,1} = 2.0$ km/s, $V_{P0,2} = 3.0$ km/s, and 
         $V_{P0,3} = 3.5$ km/s; the interval zero-offset traveltimes are 
         equal to each other ($\tau_1 = \tau_2 = \tau_3 = 1.0$ s).
         The relevant anisotropic parameters [in Tsvankin's\cite{Tsvankin1997} notation] 
         are (subscripts denote the layer number): 
         Layer 1 -- $\delta^{\rm (1)}_1 = 0.25$, $\delta^{\rm (2)}_1 = -0.15$, 
         Layer 2 -- $\delta^{\rm (1)}_2 = -0.20$, $\delta^{\rm (2)}_2 = 0.20$, 
         Layer 3 -- $\delta^{\rm (1)}_3 = 0.25$, and $\delta^{\rm (2)}_3 = -0.15$.
        }
\label{fig032}
\end{figure}

To quantify this conclusion, we consider two numerical examples. Figure~\ref{fig032} shows the azimuthally-dependent $P$-wave NMO velocity in an orthorhombic medium consisting of three horizontal layers with strong azimuthal anisotropy. While the exact NMO ellipse (solid line) happens to be close 
to a circle, the approximate, rms-averaged normal-moveout velocity (dashed line) has an oval noncircular shape because the interval NMO ellipses are far different from circles. The maximum error of the rms averaging is 
about 6.3\%, which will lead to much higher errors in the interval velocities 
after application of the Dix differentiation~(\ref{eq0265}). Evidently,
for this model it is necessary to use the exact NMO equation,
which properly accounts for the influence of azimuthal anisotropy on normal 
moveout.

For models with moderate azimuthal anisotropy and a horizontal reflector (i.e., with the interval NMO-velocity variation limited by 10-20\%), the accuracy of the rms averaging of NMO velocities is much higher. This implies that for such media it is possible to obtain the interval NMO velocity by the conventional Dix differentiation at a given azimuth. In the special case of horizontally layered HTI media (transverse isotropy with a horizontal axis of symmetry), the same conclusion was made by Al-Dajani and Tsvankin\cite{AlDajaniTsvankin1996}. 

It should be emphasized, however, that for {\it dipping}$\,$ reflectors the Dix differentiation {\it cannot}$\,$ be applied in the standard fashion (even if the rms-averaging equation provides sufficient accuracy) because the interval NMO velocities are still calculated for non-existent reflectors and cannot be found directly from the data. In the presence of anisotropy, interval parameter estimation using dipping events is impossible without a layer-stripping procedure that requires reconstruction of the NMO ellipses in the overburden and, therefore, cannot be carried out for a single azimuth. 

On the whole, we would recommend to use the generalized Dix equation for {\it any} azimuthally anisotropic model, provided the azimuthal coverage of the data is sufficient to reconstruct the dependence $V_{\rm nmo} (\alpha)$. Since our algorithm operates with the NMO ellipses rather than individual azimuthal moveout measurements, it has the additional advantage of smoothing the azimuthal variation of NMO velocity, which helps to eliminate ``outliers'' and stabilize the interval parameter estimation. A field-data application of the generalized Dix equation is discussed below.  

\begin{figure}
\includegraphics[width = 11.0 cm]{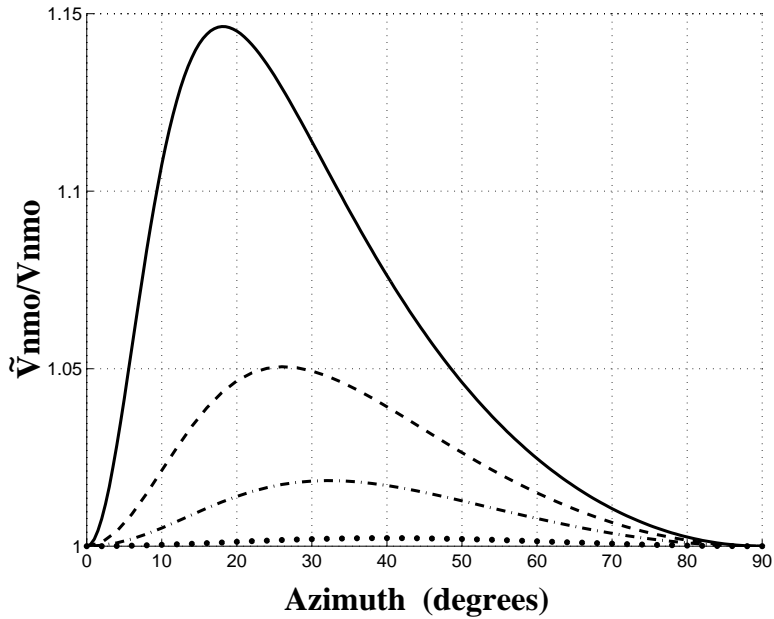} \vspace{-5mm}
%\centerline{\epsfxsize=11.0cm  \epsffile{Fig09.eps}}
\caption{The rms-averaged NMO velocity $\tilde V_{\rm nmo}$ normalized by the exact value in {\it isotropic} media; the azimuth is measured  
         with respect to the dip plane of the reflector. The model contains
         three layers above the reflector with the interval velocities 
         $V_1 = 2.0$ km/s, $V_2 = 3.0$ km/s, and $V_2 = 3.5$ km/s and
         the interval zero-offset traveltimes 
         $\tau_1 = \tau_2 = \tau_3 = 1.0$ s.
         The reflector dips are $\phi = 40^\circ$ (dotted line),
         $\phi = 60^\circ$ (dashed-dotted), 
         $\phi = 70^\circ$ (dashed),
         and $\phi = 80^\circ$ (solid).
        }
\label{fig033}
\end{figure}

Another example, in which the interval NMO ellipses differ from circles due 
to the influence of reflector dip in a purely \emph{isotropic} layered model, 
is shown in Figure~\ref{fig033}. Obviously, in this model the dip
plane of the reflector always represents a symmetry plane, and one of the 
axes of all interval NMO ellipses is parallel to the dip direction. As shown 
in the previous section, in this case the rms averaging of the interval NMO 
velocities [equations~(\ref{eq0264}) or~(\ref{eq0265})] becomes exact for 
the dip (azimuth $\alpha = 0^\circ$) and strike CMP lines (azimuth 
$\alpha = 90^\circ$), where the interval NMO values are well known\cite{Levin1971}. Figure~\ref{fig033} corroborates this conclusion: for 
azimuths $\alpha = 0^\circ$ and $\alpha = 90^\circ$ the rms-averaged NMO velocity $\tilde V_{\rm nmo}$ is equal to the exact value $V_{\rm nmo}$. In all other azimuths, 
equation~(\ref{d002}) gives only an approximation to the exact NMO velocity.
However, Figure~\ref{fig033} indicates that this approximation is quite 
accurate for small and moderate reflector dips. The maximum error of 
equation~(\ref{d002}), for example, is only 0.22\% for reflector dip 
$\phi = 40^\circ$ and 1.85\% for dip $\phi = 60^\circ$. Clearly, the error 
increases with dip because the interval NMO ellipses become more elongated 
and diverge more from a circle. 

Again, since the reflector is dipping, the interval NMO velocities in  Figure~\ref{fig033} are 
calculated for the nonzero horizontal components of the slowness vector 
determined by the reflector dip. These interval velocities correspond to 
non-existent reflectors and need to be recalculated from the NMO velocities 
of the horizontal events (which is, however, straightforward for isotropic 
media). 

%=============================================================================%

\section{Inhomogeneous anisotropic media}

\subsection{Summary of ray tracing}

Here, we give a brief overview of ray-theory equations for anisotropic media\cite{Cerveny1972, CervenyMolotkovPsencik1977, KendallThomson1989}, 
%(e.g., \v{C}erveny 1972; \v{C}erveny, Molotkov, and P\v{s}en\v{c}\'{\i}k 1977; Kendall and Thomson 1989), 
which we use below to obtain normal-moveout velocity in the presence of both anisotropy and inhomogeneity. The wave 
equation can be written in the frequency 
domain as
\begin{equation}
  \rho \omega^2 u_i  + 
  {\partial \over {\partial x_j}} \left( c_{ijkl} \, 
              {{\partial u_l} \over {\partial x_k}} \right) = 0 \, ,  
  \label{eq001}
\end{equation}
where $\omega$ is the angular frequency, $\rho \equiv \rho({\bf x})$ is
the density, $c_{ijkl} \equiv c_{ijkl}({\bf x}) = \rho({\bf x})
a_{ijkl}({\bf x})$ is the elasticity tensor in the Cartesian coordinates 
${\bf x}$, and~${\bf u} \equiv {\bf u}({\bf x})$ is the displacement vector. 
The indexes $i,j,k,l$ take on values from 1 to 3; summation over repeated 
indexes is implied.

Within the framework of ray theory, the displacement field is sought in the 
form of a series expansion,
   \begin{equation}
      {\bf u}({\bf x}, \omega) =
      \sum_{n=0}^\infty { {{\bf U}^{(n)}({\bf x})} \over {(-i \omega)^n } }
      \exp^{i \omega \tau({\bf x})} .
   \label{eq002}
   \end{equation}

Substituting this trial solution into equation~(\ref{eq001}) and
retaining only the leading (zeroth-order) term of the series~(\ref{eq002}) 
yields 
\begin{equation}
      (a_{ijkl} \, p_j p_k - \delta_{il}) A_l = 0 \, ,
  \label{eq003}
\end{equation}
where ${\bf A} \equiv {\bf U}^{(0)}$, 
$p_j \equiv p_j({\bf x}) = \partial \tau / \partial x_j$ is the slowness 
vector, $\delta_{il}$ is the symbolic Kronecker delta, and~${\bf A}$ is 
the polarization vector. Note that the slowness vector $\mathbf{p}$ is normal 
to the wavefront $\tau({\bf x}) = {\rm constant}$. From equation~(\ref{eq003}) 
it is clear that a non-trivial (non-zero) solution for the vector 
${\bf A}$ exists only if the following (Christoffel) equation is satisfied:
\begin{equation}
      F(\mathbf{p}) \equiv \det [a_{ijkl} p_j p_k - \delta_{il}] = 0 \, .
  \label{eq004}
\end{equation}

For real quantities $p_j$ corresponding to homogeneous waves, solutions 
${\bf A}^{(r)}$ ($r=1,2,3$) of equation~(\ref{eq003}) are real and orthogonal 
to each other. Therefore, they can be used to form an orthonormal basis:
\begin{equation}
      {\bf A}^{(r)} \cdot {\bf A}^{(s)} = \delta_{rs} \, .
  \label{eq006}
\end{equation}

Since $p_j = \partial \tau / \partial x_j$ depends on ${\bf x}$ in
heterogeneous media, equation (\ref{eq003}) can be regarded as a
non-linear partial differential equation for the function $\tau({\bf x})$.
The Hamilton-Jacobi theory --- or method of characteristics (Courant and
Hilbert 1962) --- can be used to rewrite this equation in the form of the
ordinary differential equations (the so-called ``ray'' equations):
\begin{eqnarray}
  \label{eq0081}
  { {d x_m} \over {d \tau} }&=&{1 \over 2} \, 
    {\partial H\over\partial{p}_m}= a_{imkl} A_i p_k A_l \, , \nonumber\\
  &&\\
  { {d p_m} \over {d \tau} }&=&-{1 \over 2} \, 
  {\partial H\over\partial{x}_m}= - {1 \over 2} \, 
  {{\partial a_{ijkl}} \over {\partial x_m} } \, A_i p_j p_k A_l \, , 
                                   \quad (m=1,2,3) \, . \nonumber
\end{eqnarray}
The Hamiltonian $H$, obtained from equation (\ref{eq003}), is given by
\begin{equation}
   H \equiv H({\bf p}, {\bf x}) = a_{ijkl} A_i p_j p_k A_l = 1 \, .
\label{hamiltonian}
\end{equation}
Note that the first of equations~(\ref{eq0081}) defines the group velocity,
\begin{equation}
  \label{eq008b}
{\bf g}\equiv{d\mathbf{x}\over{d}\tau} \, . 
\end{equation}
Substituting the first equation~(\ref{eq0081}) and equation~(\ref{eq008b})
into~(\ref{hamiltonian}), we obtain an important relation between the
slowness and the group velocity vectors
\begin{equation}
      {\bf p} \cdot {\bf g} = 1 \, .
  \label{eq008a}
\end{equation}
Since $\mathbf{p}=\mathbf{n}/V$, where ${\bf n}$ is the unit vector in the phase 
(slowness) direction, and~$V$ is the phase velocity, equation~(\ref{eq008a}) 
can be further rewritten as a relation between phase and group velocities,
\begin{equation}
  \label{eq008c}
  \mathbf{n}\cdot\mathbf{g}=V \, .
\end{equation}

For rays emanating from a point source located at the origin of the coordinate 
system, the ray-tracing equations~(\ref{eq0081}) should be supplemented 
by the following initial conditions:
\begin{equation}
  \label{eq0082}
  \mathbf{x}^{(0)}=\mathbf{0}\,,\qquad\mathbf{p}^{(0)}={\mathbf{n}^{(0)}\over V^{(0)}} \, .
\end{equation}
The ray-tracing system~(\ref{eq0081}) combined with the initial 
conditions~(\ref{eq0082}) can be solved by numerical integration using, for 
instance, the Runge-Kutta method.

%------------------------------------------------------------------------------

\subsection{Computation of NMO velocity}

The results of Grechka and Tsvankin\cite{GrechkaTsvankin1998}, briefly reviewed above, show that 
there is no need to perform a full-scale multi-azimuth ray tracing to compute 
reflection traveltimes on conventional CMP spreads. It is clear from 
equation~(\ref{eq009}) that the NMO ellipse~(\ref{b001}) and 
conventional-spread moveout as a whole are fully defined by only three 
quantities -- $W_{11}$, $W_{12}$, and $W_{22}$. Thus, three well-separated 
azimuthal measurements of $V_{\rm nmo}(\alpha)$ [which usually can be 
obtained using hyperbolic semblance analysis based on equation~(\ref{iteq:v1})] 
are sufficient to reconstruct the NMO ellipse and find the NMO velocity for 
any azimuth $\alpha$. In practice, the values of $V_{\rm nmo}(\alpha)$ 
determined on finite CMP spreads may be distorted by the influence of 
nonhyperbolic moveout. However, reflection moveout (especially that of 
$P$-waves) for spreadlengths close to the distance of the CMP from the 
reflector is typically close to hyperbolic; this has been shown in a number 
of publications\cite{TsvankinThomsen1994, Tsvankin1995, GrechkaTsvankin1998} and is further illustrated by numerical examples in this work. 
 
Although calculation of $W_{i j}$ from $V_{\rm nmo}(\alpha)$ obtained in
three azimuths is much more efficient than multi-azimuth ray tracing, it 
still requires a considerable amount of computation and does not take 
advantage of the explicit expressions for the parameters of the NMO-velocity 
ellipse discussed above. It is much more attractive to build the NMO ellipse 
directly from equations~(\ref{eq009}) and~(\ref{eq0091}), which requires 
obtaining the spatial derivatives of the ray parameter 
${\partial p_i} / {\partial x_j}$ at the CMP location (i.e., for the 
zero-offset ray). Here, we outline an efficient method of calculating these 
derivatives based on the dynamic ray-tracing equations for the zero-offset ray. 

Let us consider the zero-offset ray in the ray coordinates 
$(\gamma_1, \gamma_2, \tau)$. The parameter $\tau$ has the meaning of the 
traveltime along the ray, while $\gamma_1$ and $\gamma_2$ are supposed to 
uniquely determine the ray path and can be chosen, for instance, as the 
horizontal components of the slowness vector ($p^{}_1$ and $p_2$). Here, we use 
another option suggested by Kashtan\cite{Kashtan1982} and Kendall and Thomson\cite{KendallThomson1989},
and define $\gamma_1$ and $\gamma_2$ as the polar and azimuthal angles of 
the slowness (wave-normal) vector (respectively): 
\begin{equation}
  \label{ndef}
  \mathbf{n}=(\sin \gamma_1 \cos\gamma_2, \sin \gamma_1 \sin \gamma_2,
          \cos \gamma_1) \, .
\end{equation}

The derivatives ${\partial p_i} / {\partial x_j}$,
needed to calculate $V_{\rm nmo}(\alpha)$, can be formally written as
\begin{equation}
      {{\partial p_i} \over {\partial x_j}} = 
      {{\partial p_i} \over {\partial \gamma_1}} \,
      {{\partial \gamma_1} \over {\partial x_j}} +
      {{\partial p_i} \over {\partial \gamma_2}} \,
      {{\partial \gamma_2} \over {\partial x_j}} +
      {{\partial p_i} \over {\partial \tau}} \, 
      {{\partial \tau} \over {\partial x_j}}  \, .
  \label{eq013}
\end{equation}
Using the matrix notation
\begin{equation}
  \label{eq014}
   {\bf P} = \left[{{\partial {\bf p}} \over {\partial \gamma_1}}, \,
                   {{\partial {\bf p}} \over {\partial \gamma_2}}, \,
                   {{\partial {\bf p}} \over {\partial \tau}} \right] \, ,
\qquad
   {\bf X} = \left[{{\partial {\bf x}} \over {\partial \gamma_1}}, \,
                   {{\partial {\bf x}} \over {\partial \gamma_2}}, \,
                   {{\partial {\bf x}} \over {\partial \tau}} \right] \,
\end{equation}
and the fact that the inverse matrix ${\bf X}^{-1}$ contains the rows
\[
   {\bf X}^{-1} = \left[ 
         \begin{array}{c}
               {\partial \gamma_1} / {\partial {\bf x}} \\
               {\partial \gamma_2} / {\partial {\bf x}} \\
               {\partial \tau} / {\partial {\bf x}} 
         \end{array}  \right] \, ,
\]
we represent equation~(\ref{eq013}) in the form
\begin{equation}
      {{\partial p_i} \over {\partial x_j}} = {\bf P} \,{\bf X}^{-1} \, .
  \label{eq015}
\end{equation}
Hence, if the matrices~(\ref{eq014}) have been calculated for the zero-offset 
ray at the CMP (surface) location, the derivatives 
${\partial p_i} / {\partial x_j}$, ($i,j = 1,2$) can be determined as the 
upper-left $2 \times 2$ submatrix of the $3 \times 3$ matrix~(\ref{eq015}). 
After computing the zero-offset traveltime $\tau_0$ using kinematic ray 
tracing, we can find the NMO velocity from equations~(\ref{eq009}) 
and~(\ref{eq0091}). Note that the values of ${\partial p_i} / {\partial x_j}$ 
used in the NMO-velocity calculation correspond to one-way propagation from 
the zero-offset reflection point to the surface\cite{GrechkaTsvankin1998}. 
In other words, both {\bf p} and {\bf x} should be computed for rays emanating 
from an imaginary source located at the reflection point of the zero-offset 
ray.

The third column of the matrices {\bf P} and {\bf X} [i.e. the derivatives 
${\partial {\bf p}} / {\partial \tau}$ and 
${\partial {\bf x}} / {\partial \tau}$] can be obtained using the kinematic 
ray-tracing equations (\ref{eq0081}). To find the first and second columns 
[i.e., the derivatives ${\partial {\bf p}} / {\partial \gamma_n}$ and
${\partial {\bf x}} / {\partial \gamma_n}$, ($n = 1,2$)], let us consider the 
so-called dynamic ray-tracing equations responsible for the geometrical 
spreading along the ray\cite{CervenyMolotkovPsencik1977, KendallThomson1989}. 
%(e.g., \v{C}erveny, Molotkov and P\v{s}en\v{c}\'{\i}k 1977; Kendall and Thomson 1989). 
These equations are 
obtained by differentiating the kinematic ray-tracing system~(\ref{eq0081}) 
with respect to $\gamma_1$ and $\gamma_2$:  
\begin{eqnarray}
  \label{eq0162}
    {d \over {d \tau}} 
       \left( { {\partial x_m} \over {\partial \gamma_n} } \right)&=& 
         {\partial \over {\partial \gamma_n} } \, 
          \left( a_{imkl} A_i p_k A_l \right) , \nonumber\\
&&\\
    {d \over {d \tau}}
       \left( { {\partial p_m} \over {\partial \gamma_n} } \right)&=&
      - {1 \over 2} \, {\partial \over {\partial \gamma_n} } \, 
           \left( {{\partial a_{ijkl}} \over {\partial x_m} } \,
                  A_i p_j p_k A_l \right) , 
                  \quad (n=1,2; \,~ m=1,2,3) \, . \nonumber
\end{eqnarray}

The initial conditions for these equations are, in turn, derived by 
differentiating the corresponding initial conditions~(\ref{eq0082}) for the 
kinematic ray-tracing equations (\ref{eq0081}). Taking into account 
equation~(\ref{eq008c}), we find
\begin{eqnarray}
  \label{eq0242}
    { {\partial {\bf x}^{(0)}} \over {\partial \gamma_n} }&=&{\bf 0} \, , 
                                                              \nonumber\\
    { {\partial {\bf p}^{(0)}} \over {\partial \gamma_n} }&=& 
    { 1 \over V^{(0)}} \left[ { {\partial {\bf n}^{(0)}} \over 
                                {\partial \gamma_n} } -
                        { {\bf n}^{(0)} \over V^{(0)} }
                        \left( {\bf g}^{(0)} \cdot 
                        { {\partial {\bf n}^{(0)}} \over {\partial \gamma_n} }
                        \right) \right] , \quad (n=1,2) \, ,
\end{eqnarray}
where $V^{(0)}$, ${\bf n}^{(0)}$, and ${\bf g}^{(0)}$ are the phase velocity,
the unit vector in the wave-normal (phase) direction and the group-velocity 
vector at the source location. In our case, the velocities $V^{(0)}$ and 
${\bf g}^{(0)}$ should be evaluated immediately above the reflector at the 
zero-offset reflection point (the effective source). The derivatives of the 
wave-normal vector ${\partial {\bf n}^{(0)}}/{\partial \gamma_n}$ 
can be computed in a straightforward way from equation~(\ref{ndef}).
 
Thus, the derivatives needed to obtain the normal-moveout velocity are 
exactly the same as those required to compute the geometrical spreading along the zero-offset ray. This result is not entirely surprising because NMO velocity is related to the wavefront curvature\cite{Shah1973}, which, in turn, determines geometrical spreading. Therefore, the azimuthally-dependent NMO velocity in inhomogeneous arbitrary anisotropic media can be computed by integrating the dynamic ray-tracing 
equations~(\ref{eq0162}) for the one-way zero-offset ray and substituting 
the results into equations~(\ref{eq015}), ~(\ref{eq0091}) and~(\ref{eq009}). 
Since this approach requires tracing of only \emph{one} zero-offset ray 
together with the derivatives~(\ref{eq0162}), it is orders of magnitude less 
time consuming than is the tracing of hundreds of reflected rays for 
different azimuths and source-receiver offsets as would otherwise be required. 
Moreover, as shown in the next section, our algorithm becomes significantly 
simpler, both analytically and computationally, if the model consists of 
homogeneous layers or blocks.

%------------------------------------------------------------------------------

\subsection{Piecewise homogeneous media} 

Let us consider a medium composed of arbitrary anisotropic
{\it homogeneous} layers (or blocks) separated by smooth interfaces. In this 
case, the ray trajectory is piecewise linear, and integration of the kinematic 
ray-tracing equations~(\ref{eq0081}) reduces to summation along straight ray 
segments:
\begin{eqnarray}
  \label{eq0172}
  {\bf x}^{(\ell)}&=&{\bf x}^{(\ell-1)} + {\bf g}^{(\ell)} \, \tau^{(\ell)} \,,
  \nonumber\\&&\\
      {\bf p}^{(\ell)}&=&{\rm const} \, ,\nonumber
\end{eqnarray}
where ${\bf x}^{(\ell-1)}$ denotes the ray coordinate at the interface
between the $\ell-1$-th and $\ell$-th layer, $\tau^{(\ell)}$ is the traveltime
inside the $\ell$-th layer, and $g^{(\ell)}$ is the group velocity in this
layer. Differentiation of equations~(\ref{eq0172}) yields two derivatives 
required in the computation of the NMO ellipse: 
\begin{equation}
      {\partial {\bf x}} / {\partial \tau} = {\bf g} \, ,
  \label{eq0782}
\end{equation}
\begin{equation}
      {\partial {\bf p}} / {\partial \tau} = 0 \, ,
  \label{eq0783}
\end{equation}
with the group-velocity vector {\bf g} evaluated for the zero-offset ray at 
the CMP location.

To fully describe the ray path (for purposes of kinematic ray tracing), 
equations~(\ref{eq0172}) must be supplemented by the boundary conditions at 
model interfaces for {\bf x} and {\bf p}. The boundary conditions will also 
be used in the equations for dynamic ray tracing discussed below. Since the 
ray has to be continuous,
\begin{equation}
      {\bf x}^{(\ell)} = {\bf x}^{(\ell-1)} \, .
  \label{eq0181}
\end{equation}
The ray parameter {\bf p} satisfies Snell's law,
\begin{equation}
      {\bf p}^{(\ell)} \times {\bf b}^{(\ell)} = 
      {\bf p}^{(\ell-1)} \times {\bf b}^{(\ell)} \, ,
  \label{eq0182}
\end{equation}
where ``$\times$'' denotes the cross product and ${\bf b}^{(\ell)}$ is
the unit vector normal to the $\ell$-th interface at ${\bf x}^{(\ell)}$.

Integration of the dynamic ray-tracing equations (\ref{eq0162}) in the case
of homogeneous layers is relatively straightforward as well. Continuation of 
the derivatives 
${\partial {\bf x}} / {\partial \gamma_n}$ and 
${\partial {\bf p}} / {\partial \gamma_n}$ across the $\ell$-th layer is
expressed by
\begin{equation}
      { {\partial {\bf x}^{(\ell+1)}} \over {\partial \gamma_n} } =
      { {\partial {\bf x}^{(\ell)}} \over {\partial \gamma_n} } +
      { {\partial {\bf g}^{(\ell+1)}} \over {\partial \gamma_n} } \, 
      \tau^{(\ell+1)}
  \label{eq0191}
\end{equation}
and
\begin{equation}
      { {\partial {\bf p}^{(\ell+1)}} \over {\partial \gamma_n} } =
      { {\partial {\bf p}^{(\ell)}} \over {\partial \gamma_n} } \,,
                                         \quad (n = 1, 2) \, .
  \label{eq0192}
\end{equation}
The derivative of the group velocity needed in equation~(\ref{eq0191}) is 
obtained in Appendix~D from equations~(\ref{eq008b}) and~(\ref{eq0081}).

To propagate the derivatives 
${\partial {\bf x}} / {\partial \gamma_n}$ and 
${\partial {\bf p}} / {\partial \gamma_n}$ across the $\ell$-th (smooth) 
interface, Kashtan\cite{Kashtan1982} suggested differentiating equations~(\ref{eq0181}) 
and Snell's law [equation~(\ref{eq0182})] with respect to $\gamma_n$. His 
results become especially simple for a plane interface with the normal 
${\bf b}^{(\ell)}$:
\begin{equation}
  { {\partial {\bf x}^{(\ell)}} \over {\partial \gamma_n} } =
  { {\partial {\bf x}^{(\ell-1)}} \over {\partial \gamma_n} } +
  { {\bf g}^{(\ell)} - {\bf g}^{(\ell-1)} \over 
    {\bf g}^{(\ell-1)} \cdot {\bf b}^{(\ell)}} \:
  \left( {\bf b}^{(\ell)} \cdot
         {\partial {\bf x}^{(\ell-1)} \over {\partial \gamma_n}} \right)
  \label{eq0231}
\end{equation}
and
\begin{equation}
  { {\partial {\bf p}^{(\ell)}} \over {\partial \gamma_n} } =
  { {\partial {\bf p}^{(\ell-1)}} \over {\partial \gamma_n} } -
  {{\bf b}^{(\ell)}\over {\bf b}^{(\ell)} \cdot {\bf g}^{(\ell)}} \:
  \left({\bf g}^{(\ell)}\cdot 
  {\partial {\bf p}^{(\ell-1)} \over {\partial \gamma_n}} \right) \,, 
                               \quad (n=1,2) \,.
  \label{eq0232}
\end{equation}

Equations~(\ref{eq0191}) -- (\ref{eq0232})
allow us to continue the initial values of the derivatives 
${\partial {\bf x}} / {\partial \gamma_n}$ and
${\partial {\bf p}} / {\partial \gamma_n}$ [equations~(\ref{eq0242})] from 
the zero-offset reflection point through our layered (or blocked) model to 
the surface. Since the quantities needed to obtain these derivatives (i.e., 
the group velocity and traveltime) have to be found for the kinematic 
ray-tracing anyway, the additional computational cost of this operation is 
minimal. Finally, we calculate the derivatives 
${\partial p_i} / {\partial x_j}$ from equation~(\ref{eq015}) and construct 
the $V_{\rm nmo}(\alpha)$ ellipse defined by equations~(\ref{eq009}) 
and~(\ref{eq0091}). 

\section{Synthetic examples for inhomogeneous media} 

The accuracy of our single-layer NMO equation has been discussed above (see Figure~\ref{fig04}). Here, we carry out synthetic tests to compare the hyperbolic moveout equation parameterized by the exact NMO velocity with ray-traced traveltimes for inhomogeneous anisotropic models.
 
\begin{figure}
\includegraphics[width = 3.1 in]{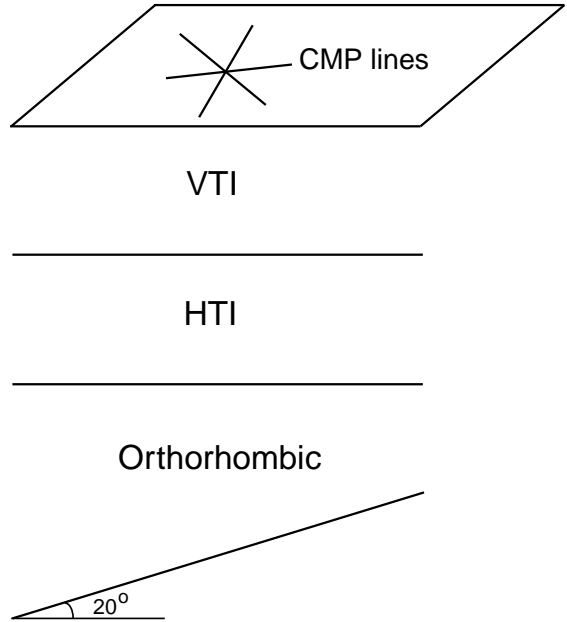} \vspace{-5mm}
\caption{The model used in Figure~\protect\ref{test} to check the accuracy of the generalized Dix equation. Layer 1 is transversely isotropic with a vertical symmetry axis (VTI) and relevant parameters $V_{P0,1}=2.5$ km/s, $\epsilon_1=0.2$, $\delta_1=0.1$. 
  Layer 2 is TI with a horizontal symmetry axis (azimuth $\beta_2=30^\circ$) and  $V_{P0,2}=3.0$ km/s,
         $\epsilon^{\rm (V)}_2=-0.2$, $\delta^{\rm (V)}_2=-0.15$ (for HTI notation, see Tsvankin\cite{TsvankinHTI1997}). Layer 3 is orthorhombic with  $V_{P0,3}=3.5$ km/s, $\epsilon^{\rm (1)}_3=0.2$, $\delta^{\rm (1)}_3=0.15$,
         $\epsilon^{\rm (2)}_3=-0.3$, $\delta^{\rm (2)}_3=-0.2$,
         $\delta^{\rm (3)}_3=0.05$; the azimuth of the $[x_1,x_3]$
         symmetry plane $\beta_3=60^\circ$. 
         The interface depths are $z_1 = 1.0$ km, $z_2 = 2.0$ km,
         $z_3 = 3.0$ km.  The reflector dip is $20^\circ$, the azimuth of the dip plane is $0^\circ$. 
}
\label{modeldix}
\end{figure}

\begin{figure}
\includegraphics[width = 3.8 in]{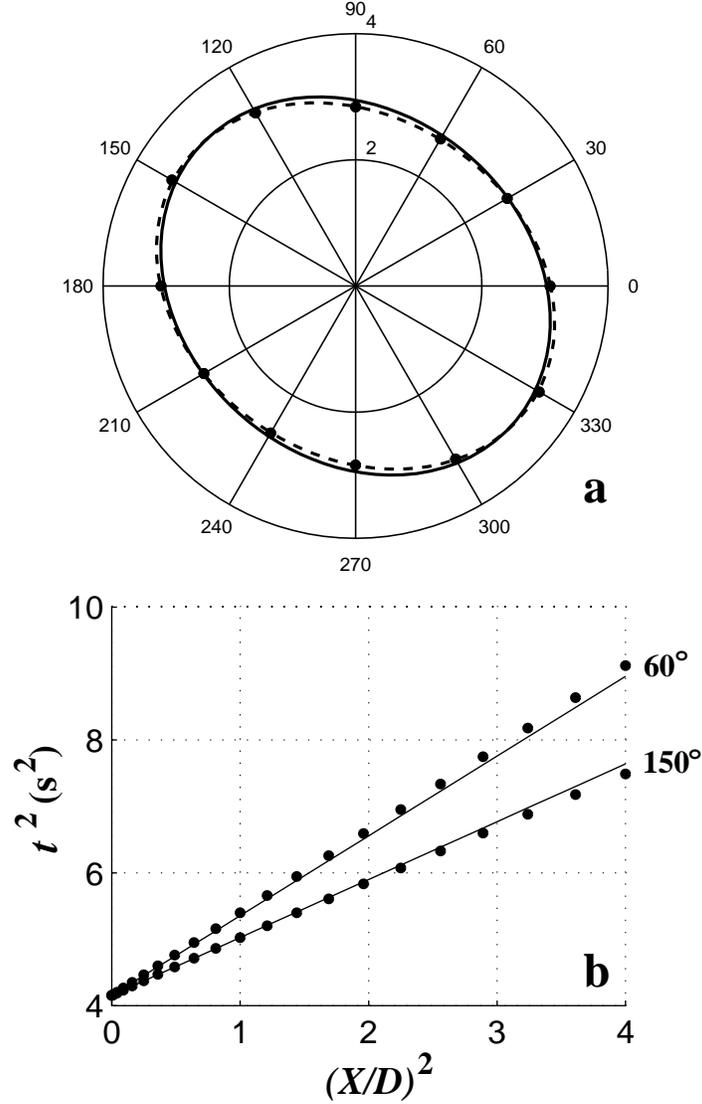} \vspace{-9mm}
\caption{ (a) Comparison between the theoretical $P$-wave NMO ellipse calculated from the generalized Dix equation (solid) and moveout velocities obtained from ray-traced traveltimes for spreadlength equal to the distance between the CMP and the reflector (dots). The model is shown in Figure~\protect\ref{modeldix}; the dashed line marks the best-fit ellipse found from the finite-spread moveout velocities. (b) Hyperbolic moveout curve parameterized by the exact NMO velocity (solid) vs. computed traveltimes (dots) at azimuths $60^\circ$ and $150^\circ$; $D$ is the CMP-reflector distance.
   }
\label{test}
\end{figure}

Figure~\ref{test} illustrates the performance of the Dix
equation~(\ref{dixavr}) for a model that includes three anisotropic layers with different symmetry above a dipping reflector (Figure~\ref{modeldix}). We used ray tracing 
to calculate $P$-wave reflection traveltimes along six azimuths with 
increment $30^\circ$ and obtained moveout velocities (dots in Figure~\ref{test}a) by fitting a hyperbola to the exact moveout. Despite the complexity of the model, the best-fit ellipse found from the finite-spread moveout velocities (dashed) are sufficiently close to the theoretical NMO ellipse (solid) computed from equations~(\ref{dixavr}) 
and~(\ref{eq009}). A small difference between the ellipses is caused by 
nonhyperbolic moveout associated with both anisotropy and vertical 
inhomogeneity. It is clear from Figure~\ref{test}b that the influence of nonhyperbolic moveout becomes substantial only at source-receiver offsets that exceed the distance between the CMP and the reflector.

\begin{figure}
\includegraphics[width = 3.2 in]{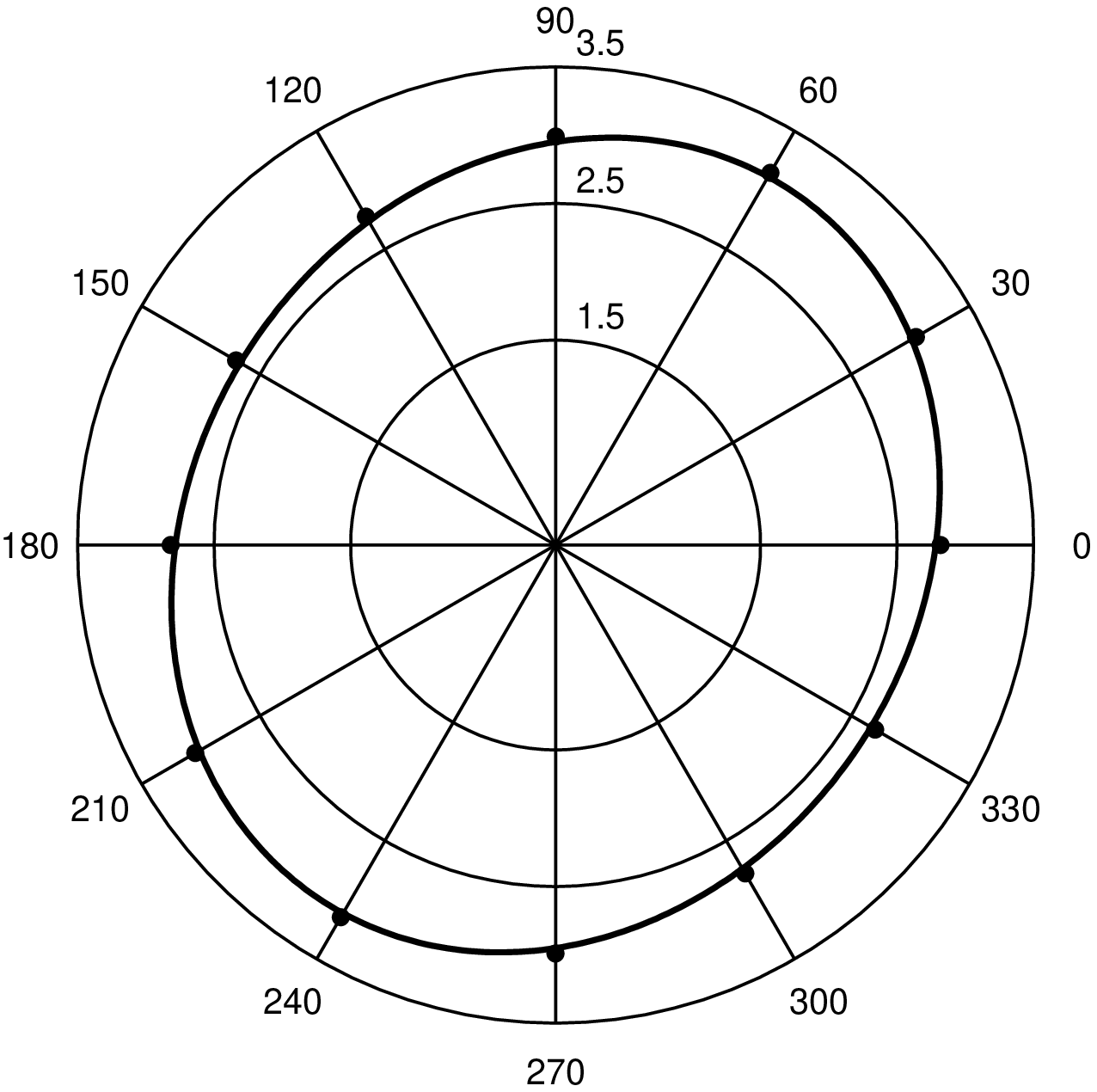} \vspace{-7mm}
%\centerline{\epsfxsize=3.6in  \epsffile{Fig03.eps}}
\caption{Comparison of the theoretical $P$-wave NMO ellipse (solid) and finite-spread moveout 
         velocity (dots; the spreadlength is equal to the CMP-reflector distance) in an azimuthally-anisotropic model with dipping layers. The NMO ellipse is computed from equations~(\protect\ref{eq009}) 
         and~(\protect\ref{eq0091}), with the spatial derivatives of the ray 
         parameter evaluated using equations~(\protect\ref{eq015}), 
         (\protect\ref{eq0782}), (\protect\ref{eq0783}), 
         and~(\protect\ref{eq0191}) through (\protect\ref{eq0232}). The model 
         consists of three dipping transversely layers with different orientation of the symmetry axis. The first layer is VTI with $V_{P0,1} = 2.0$ km/s, $\epsilon_1 = 0.2$, and $\delta_1 = 0.1$. The second layer is HTI with the azimuth of the symmetry axis of 30$^\circ$ and $V_{P0,2} = 2.4$ km/s, $\epsilon_2 = 0.15$,  $\delta_2 = 0$. The third layer is TI with a tilted symmetry axis (the azimuth is 60$^\circ$, the tilt is 30$^\circ$) and $V_{P0,3} = 3$ km/s, $ \epsilon_3 = 0.25$,  $\delta_3 = 0.08$ [for all layers we used the generic Thomsen's\cite{Thomsen1986} parameters]. The azimuth $\psi$ and dip $\phi$ of the bottom of the first layer are $\psi_1 = 70^\circ$ and $\phi_1 = 10^\circ$; for the bottom of the second layer $\psi_2 = 20^\circ$ and $\phi_2 = 15^\circ$; for the reflector $\psi_3 = 50^\circ$  and $\phi_3 = 35^\circ$. The distances between the CMP and the interfaces are 1 km, 
2 km, and 3 km.
        }
\label{fig03}
\end{figure}

A similar example, but this time for a horizontally inhomogeneous medium above the reflector is shown in Figure~\ref{fig03}. The model contains three transversely isotropic layers 
with dipping lower boundaries and differently oriented symmetry axes. The NMO ellipse (solid) provides an excellent approximation to the effective moveout velocity (dots)
for all four azimuths used in the computation, with a maximum error of 
just about 1.4\%. In addition to verifying the accuracy of our algorithm 
based on the evaluation of the derivatives 
${\partial {\bf x}} / {\partial \gamma_n}$ and 
${\partial {\bf p}} / {\partial \gamma_n}$, this test demonstrates again that the 
analytic (zero-spread) normal-moveout velocity typically provides a good 
approximation for $P$-wave reflection traveltimes on conventional spreads.

\section{Field-data example}

We applied the generalized Dix equation to a 3-D data set acquired by ARCO (with funding from the Gas Research Institute) in the Powder River Basin, Wyoming. A detailed description of this survey and preliminary processing results can be found in Corrigan et al.\cite{Corriganetal1996} and Withers and Corrigan\cite{WithersCorrigan1997}. The main goal of the experiment was to use the azimuthal dependence of $P$-wave signatures in characterization of a fractured reservoir. Hence, the acquisition was carefully designed to provide a good offset coverage in a wide range of source-receiver azimuths. To enhance the signal-to-noise ratio, the data were collected into a number of ``superbins,'' each with an almost random distribution of azimuths and offsets.

\begin{figure}
\includegraphics[width = 4.0 in]{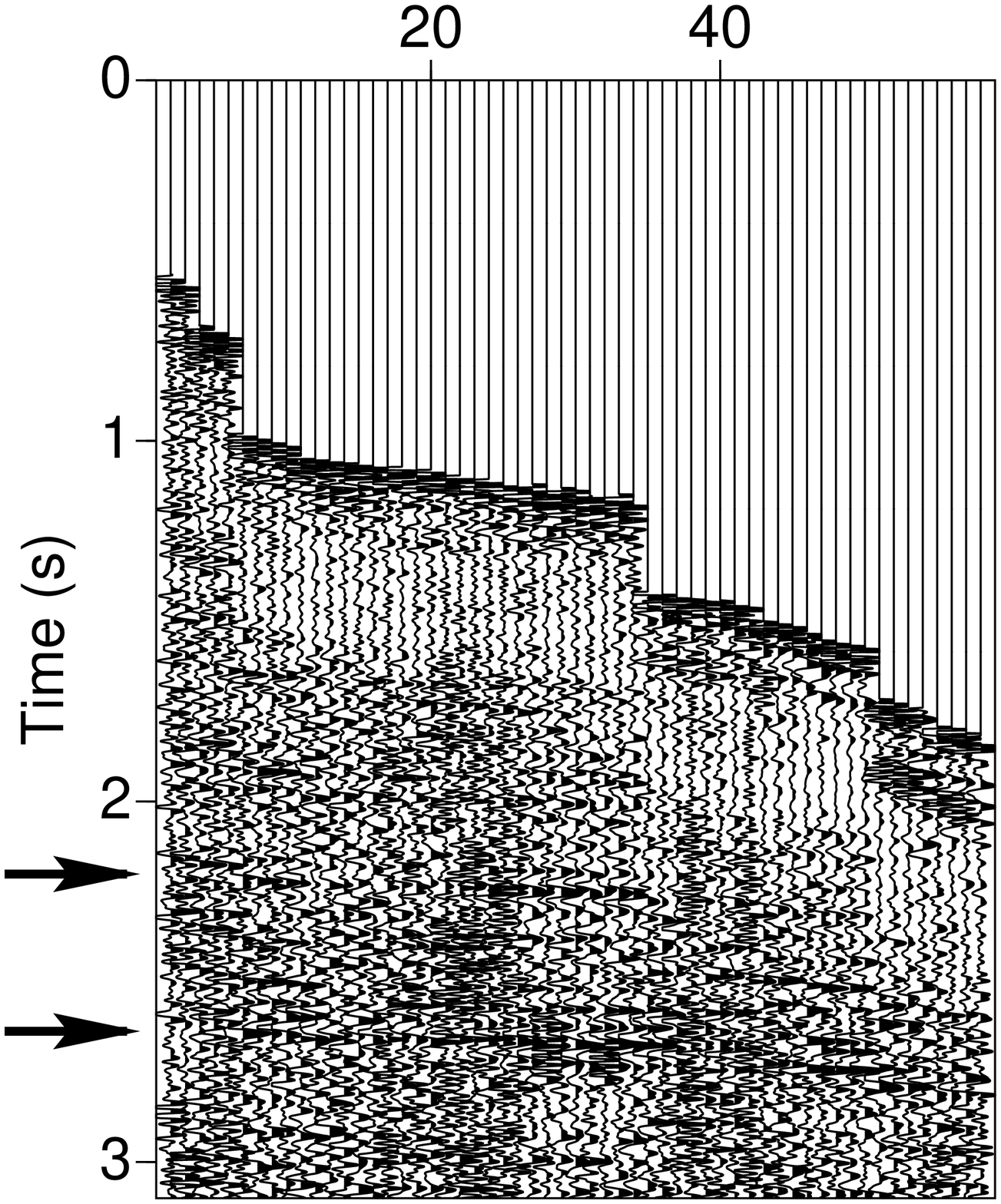} \vspace{-13mm}
%\centerline{\epsfxsize=\hsize \epsffile{Traces.eps}}
\caption{Common-midpoint gather composed of traces with source-receiver azimuths within a 20$^{\circ}$ azimuthal sector centered at N30E. 
        }
\label{traces}
\end{figure}

\begin{figure}
\includegraphics[width = 4.7 in]{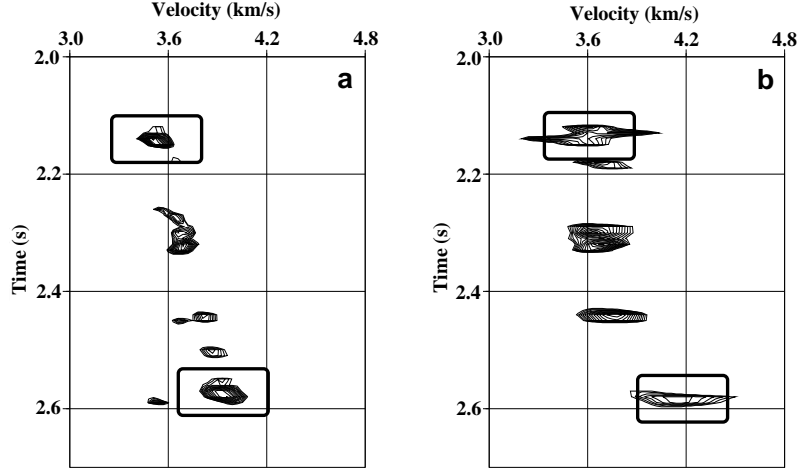} \vspace{-7mm}
%\centerline{\epsfxsize=6.7in \epsffile{fig19.eps}}
\caption{Semblance velocity panels for two azimuthal sectors centered at N130E (a) and N30E (b). The semblance maxima corresponding to the events at vertical times of 2.14 s and 2.58 s are framed.}
\label{fig19}
\end{figure}

\begin{figure}
\includegraphics[width = 6.0 in]{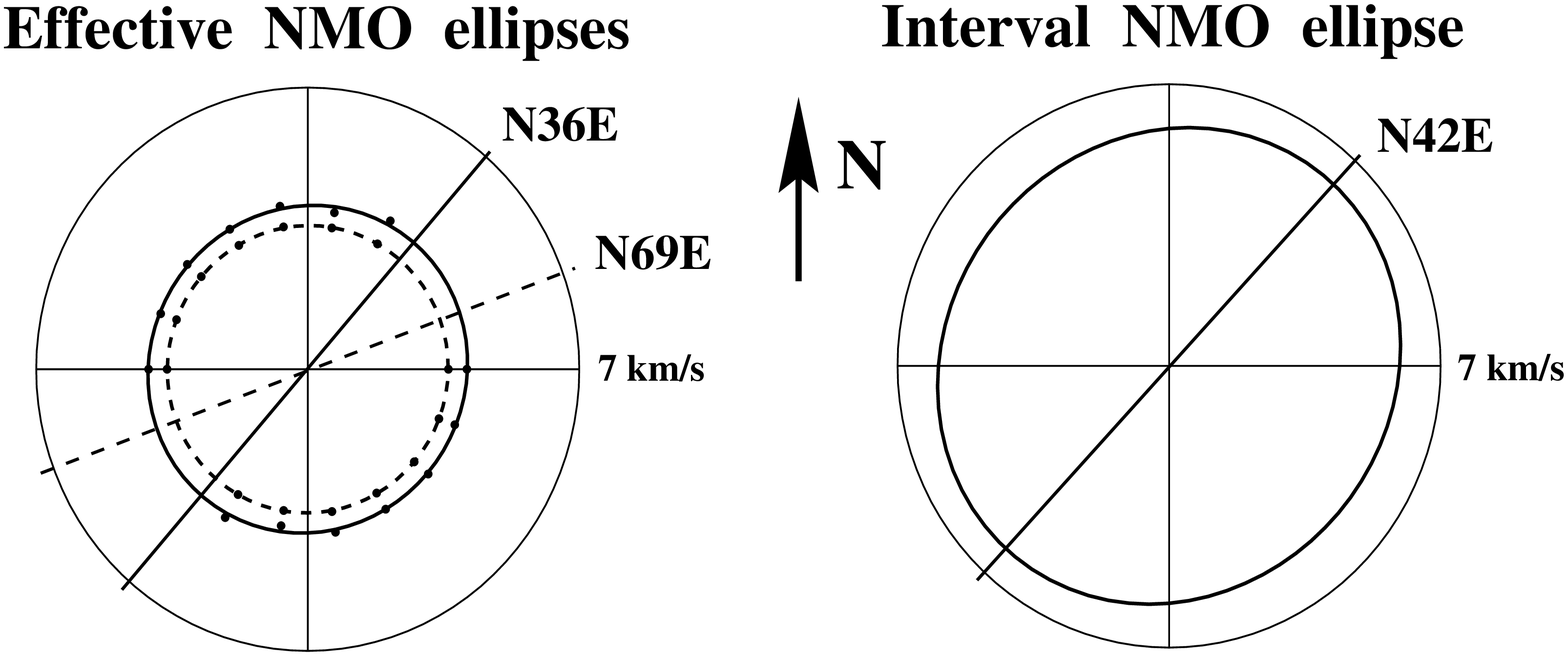} \vspace{-4mm}
%\centerline{\epsfxsize=\hsize \epsffile{fig20.eps}}
\caption{
The effective NMO ellipses for the reflection events at 2.14 s (dashed) and 2.58 s (solid) reconstructed from the data and the corresponding interval NMO ellipse obtained by the generalized Dix differentiation. The orientation of the larger semi-axis of each ellipse is marked by a radial line. 
        }
\label{fig20}
\end{figure}

Below we show the results of our velocity analysis for one of the superbins located in the southwest corner of the survey area. To obtain the azimuthal dependence of NMO velocity, we divided the traces into nine 20$^{\circ}$ azimuthal sectors and carried out conventional hyperbolic semblance analysis separately for each sector. Figure~\ref{traces} shows the composite CMP gather in one of the sectors with two prominent reflection events marked by arrows. According to Withers and Corrigan\cite{WithersCorrigan1997}, the reflection at a two-way vertical time of 2.14 s corresponds to the bottom of the Frontier/Niobrara formations, and the event at 2.58 s is the basement reflection. Semblance panels for two sectors 100$^{\circ}$ apart are displayed in Figure~\ref{fig19}. While the best-fit stacking velocity for the event at 2.14 s is weakly dependent on azimuth, the velocity for the basement reflection is noticeably higher at azimuth N30E. This observation is confirmed by the shape of the effective NMO ellipses reconstructed from the semblance panels (Figure~\ref{fig20}, left plot). The stacking velocity of the basement reflection along the larger semi-axis of the ellipse is 4\% higher than in the orthogonal direction; the corresponding number for the shallower reflection is 1.7\%. The orientation of both effective ellipses agrees with the results of Withers and Corrigan\cite{WithersCorrigan1997}, who used a different algorithm. It should be mentioned that ignoring azimuthal velocity variations on the order of 3-4\% and mixing up all source-receiver azimuths (as is conventionally done in 3-D processing) would inevitably lead to poor stacking and deterioration of the final seismic image.

Since the dips in the survey area are extremely small\cite{Corriganetal1996}, the azimuthal dependence of stacking velocity can be attributed to the influence of azimuthal anisotropy associated with vertical fractures. To study the {\it interval} properties for vertical times between 2.14 and 2.58 s, we applied the generalized Dix equation~(\ref{eq0260}) to the effective NMO ellipses. Note that the different orientation of the effective ellipses in Figure~\ref{fig20}, indicative of depth-varying principal directions of the azimuthal anisotropy, does not pose a problem for the generalized Dix differentiation. The pronounced azimuthal variation in the interval NMO velocity (9\%, Figure~\ref{fig20}) can be explained by the intense fracturing in the layer immediately above the basement. The direction of the larger semi-axis of the interval NMO ellipse is in general agreement with the predominant fracture orientation in the deeper part of the section determined from borehole data and shear-wave splitting analysis\cite{WithersCorrigan1997}. Complete processing/inversion results for the survey area will be reported in forthcoming publications.

\section{Discussion and conclusions} 

Azimuthally-dependent normal-moveout velocity 
around a certain CMP location is described by a simple quadratic form and usually has an {\it elliptical} shape, with the 
orientation and semi-axes of the ellipse determined by the properties of the 
medium and the direction of the reflector normal at the zero-offset reflection 
point. Using this general result obtained by Grechka and Tsvankin\cite{GrechkaTsvankin1998}, we 
have presented a series of solutions for the exact normal-moveout velocity of 
pure modes in anisotropic models of various complexity. For a homogeneous 
anisotropic layer above a dipping reflector, NMO velocity was found explicitly 
as a function of the slowness vector corresponding to the zero-offset ray. 
This single-layer equation is valid for arbitrary anisotropic symmetry and 
any orientation of the CMP line with respect to the reflector strike. The 
vertical component of the slowness vector and its derivatives with respect to 
the horizontal slownesses, needed to compute the NMO velocity, can be obtained 
in an explicit form using the  Christoffel equation. In addition to simplifying moveout modeling, our NMO equation can be effectively used in moveout inversion, as well as in developing weak-anisotropy approximations for different symmetries. 

If the model contains a stack of homogeneous arbitrary anisotropic layers 
above a dipping reflector, the NMO ellipse should be obtained by a Dix-type 
averaging of the single-layer expressions described above. Instead of the 
squared NMO velocities in the conventional Dix formula, our generalized 
equation operates with the interval matrices ${\bf W}_\ell$ that describe 
the NMO {\it ellipses}$\,$ corresponding to the individual layers. To 
find azimuthally-dependent normal-moveout velocity, it is sufficient to 
compute the zero-offset traveltime and the interval NMO ellipses for the 
slowness vector of the zero-offset ray. The generalized Dix equation can be 
used to perform moveout-based interval parameter 
estimation in vertically inhomogeneous anisotropic models of any symmetry. It should be emphasized, however, that application of the generalized Dix differentiation to dipping events entails full-scale layer-stripping because NMO ellipses in the individual layers cannot be directly measured from the data. 

One important special case considered in detail is a model with the same 
(throughgoing) vertical symmetry plane in all layers that also coincides 
with the dip plane of the reflector (e.g., the medium above the reflector 
is TI with a vertical symmetry axis). Because of the mirror symmetry with respect to the dip plane, the axes of the NMO ellipse are aligned with the dip and strike directions of the reflector. The generalized Dix equation in such a model reduces to the rms averaging of the dip-line and strike-line NMO velocities 
in the individual layers (these averages determine the semi-axes of the NMO 
ellipse). This result represents a 3-D extension of the Dix-type equation 
developed by Alkhalifah and Tsvankin\cite{AlkhalifahTsvankin1995} for normal moveout in the dip 
plane of the reflector. 

Except for this special case, the effective 
NMO velocity computed by the Dix rms averaging generally takes an oval {\it anelliptic}$\,$ form that thus deviates 
from the exact NMO ellipse. Still, this deviation is not significant if 
the interval NMO ellipses are close to being circles, which implies the 
absence of large dips and significant azimuthal anisotropy. In any case, it is preferable to apply the generalized Dix equation (as opposed to the conventional Dix differentiation at any given azimuth) for {\it any} azimuthally anisotropic model because in addition to being more accurate it also provides the important advantage of smoothing the effective moveout velocities using the correct (elliptical) functional form and thus reducing the instability in interval parameter estimation.

We complete the analysis by considering the most general inhomogeneous media and presenting an 
algorithm that leads to a dramatic reduction in the amount of computations 
needed to obtain the NMO velocity and conventional-spread reflection moveout. 
All information required to construct the NMO ellipse is contained in the 
results of the {\it dynamic} ray tracing (i.e., computation of geometrical spreading) of a single (zero-offset) ray. 
Although evaluation of geometrical spreading requires solving an additional system of differential 
equations together with the kinematic ray-tracing equations, this algorithm 
is orders of magnitude more efficient than  multi-offset, multi-azimuth 
ray tracing. Furthermore, if the model consists of homogeneous layers or 
blocks separated by smooth interfaces, all quantities needed to find the 
NMO ellipse can be computed during the kinematic tracing of the zero-offset 
ray. 

The normal-moveout velocity discussed here is defined in the zero-spread 
limit and cannot account for nonhyperbolic moveout caused by anisotropy and 
inhomogeneity on finite-spread CMP gathers. Nevertheless, our numerical 
examples for various anisotropic models demonstrate that the hyperbolic 
moveout equation parameterized by NMO velocity provides good accuracy 
in the description of reflection moveout (especially that of $P$-waves) on 
conventional spreads close to the distance between the CMP and the reflector. Even if the hyperbolic moveout approximation becomes inadequate, NMO velocity can be obtained by means of nonhyperbolic moveout analysis\cite{TsvankinThomsen1994, SayersSEG1995}. Hence, the results of this work can be efficiently used in traveltime inversion and dip-moveout processing for arbitrary anisotropic media. 

To show the feasibility of applying our formalism in fracture detection, we processed wide-azimuth 3-D $P$-wave data acquired over a fractured reservoir in the Powder River Basin, Wyoming. The generalized Dix differentiation allowed us to obtain the {\it depth-varying}$\,$ fracture orientation and estimate the magnitude of azimuthal anisotropy (measured by $P$-wave moveout velocity). The direction of the larger semi-axis of the interval NMO ellipse in the strongly anisotropic layer above the basement is in agreement with the fracture trend in this part of the section. Therefore, if the formation of interest has a sufficient thickness, azimuthal moveout analysis of $P$-wave (and, if available, shear-wave) data by means of the generalized Dix equation provides valuable information for characterization of fracture networks.  

\section{Acknowledgments}

We are grateful to Dennis Corrigan and Robert Withers of ARCO for providing the field data and sharing their knowledge of the processing and interpretation issues. We would like to thank Bruce Mattocks and Bob Benson (both of CSM) for helping us in data processing and Ken Larner (CSM) and Andreas R\"{u}ger (CSM; now at Landmark) for their reviews of the paper. The support for this work was provided by the members of the Consortium Project on Seismic Inverse Methods for Complex Structures at Center for Wave Phenomena, Colorado School of Mines and by the United  States Department of Energy (project ``Velocity Analysis, Parameter  Estimation, and Constraints on Lithology for Transversely Isotropic  
Sediments'' within the framework of the Advanced Computational  
Technology Initiative).

\appendix
\section{Relation between the matrix W and the NMO-velocity ellipse} 

Azimuthally dependent normal-moveout velocity is described by 
equation~(\ref{eq009}) of the main text as a general second-order curve in 
the horizontal plane. The expression for $V_{\rm nmo} (\alpha)$ can be 
simplified further by aligning the horizontal coordinate axes with the 
eigenvectors of the symmetric matrix ${\bf W}$\cite{GrechkaTsvankin1998}. 
This rotation reduces equation~(\ref{eq009}) to 
\begin{equation}
      V_{\rm nmo}^{-2} (\alpha) = \lambda_1 \cos^2 (\alpha - \beta) +
                                  \lambda_2 \sin^2 (\alpha - \beta) \, ,
  \label{b00111}
\end{equation}
where $\lambda_1$ and $\lambda_2$ are the eigenvalues of the matrix 
${\bf W}$, and $\beta$ is the angle between the eigenvector corresponding 
to $\lambda_1$ and the $x_1$-axis. 

To verify the equivalence between equations~(\ref{b00111}) and~(\ref{eq009}), 
we expand 
\[
    \cos^2 (\alpha - \beta) = \cos^2 \alpha \cos^2 \beta
              + 2 \sin \alpha \sin \beta \cos \alpha \cos \beta
                            + \sin^2 \alpha \sin^2 \beta
\]
and
\[
    \sin^2 (\alpha - \beta) = \cos^2 \alpha \sin^2 \beta
              - 2 \sin \alpha \sin \beta \cos \alpha \cos \beta
                            + \sin^2 \alpha \cos^2 \beta \, .
\]
Equations~(\ref{b00111}) and~(\ref{eq009}) are identical if
\begin{equation}
      W_{11} = \lambda_1 \cos^2 \beta + \lambda_2 \sin^2 \beta \, ,
  \label{b002}
\end{equation}
\begin{equation}
      W_{12} = {1 \over 2} \, (\lambda_1 - \lambda_2) \, \sin 2 \beta \, ,
  \label{b003}
\end{equation}
and
\begin{equation}
      W_{22} = \lambda_1 \sin^2 \beta + \lambda_2 \cos^2 \beta \, .
  \label{b004}
\end{equation}

Inverting equations (\ref{b002}) -- (\ref{b004}) for $\lambda_{1,2}$ and 
$\beta$ yields
\begin{equation}
      \lambda_{1,2} = {1 \over 2} 
             \left[W_{11} + W_{22} \pm
                   \sqrt{ (W_{11} - W_{22})^2 + 4 W_{12}^2 } \,
             \right] 
  \label{b005}
\end{equation}
and
\begin{equation}
  \tan \beta = \frac{W_{22} - W_{11}  + 
                     \sqrt{ (W_{22} - W_{11})^2 + 4 W_{12}^2 }} 
                    {2 W_{12}} \, , \quad (W_{12} \ne 0) \, .
  \label{b006}
\end{equation}
Equations~(\ref{b005}) and~(\ref{b006}) show that $\lambda_{1,2}$ are indeed 
the eigenvalues of ${\bf W}$ and $\tan\beta$ is equal to the ratio of the 
components ``2'' and ``1'' of the eigenvector corresponding to $\lambda_1$. 
If $W_{12}=0$, the matrix {\bf W} is diagonal, and equation~(\ref{eq009}) 
reduces to equation~(\ref{b00111}) without any rotation. 

As follows from equation~(\ref{b00111}), $V_{\rm nmo}(\alpha)$ represents an 
ellipse in the horizontal plane if the eigenvalues $\lambda_{1,2}$ are 
positive\cite{GrechkaTsvankin1998}. The ``principal'' values of the 
azimuthally dependent NMO velocity (the semi-axes of the ellipse) are given by
\begin{equation}
      V_{\rm nmo}^{(i)} = {1 \over {\sqrt{\lambda_i}} }, \quad (i=1,2) \, .  
  \label{b0011}
\end{equation}

%=============================================================================%

\section{NMO velocity in a single layer}

Here, we obtain the exact expression for the NMO velocity from a dipping 
reflector beneath a homogeneous arbitrary anisotropic layer. The derivation 
is based on the general equations~(\ref{eq009}) and~(\ref{eq0091}) describing 
the NMO ellipse and follows the approach suggested for the 2-D case by 
Cohen\cite{Cohen1998}.

To evaluate the derivatives ${\partial x_i}/{\partial p_j}$, we have to 
relate the horizontal ray displacements $(x_1, \, x_2)$ between the 
zero-offset reflection point and the surface to the horizontal components of 
the slowness vector $(p^{}_1, \, p^{}_2)$. We start by introducing the 
group-velocity vector {\bf g},
\begin{equation}
      x_i = g_i \tau \, ,  \quad (i = 1,2,3) \, ,
  \label{eq026}
\end{equation}
where $\tau$ is the one-way traveltime. Using the fact that the projection 
of the group-velocity vector on the slowness direction is equal to phase 
velocity [e.g., equation~(\ref{eq008a})], we can write 
\begin{equation}
   {\bf p} \cdot {\bf g} = p^{}_1 g^{}_1 + p^{}_2 g^{}_2 + p^{}_3 g^{}_3 = 1 \, .
  \label{eq027}
\end{equation}
Differentiating equation~(\ref{eq027}) with respect to $p^{}_i$ ($i = 1, 2$) 
and taking into account that the vertical slowness component $p^{}_3$ can be 
considered as a function of $p^{}_1$ and $p^{}_2$ yields 
\[
      g^{}_i = - {\partial p^{}_3 \over \partial p^{}_i} g^{}_3 - 
              {\bf p} \cdot {\partial {\bf g} \over \partial p^{}_i} \, , 
                           \quad (i = 1,2) \, .
 \]

Since the slowness vector ${\bf p}$ is normal to the group-velocity surface 
(wavefront) ${\bf g} (p^{}_1, \, p^{}_2)$, while the vectors 
${\partial {\bf g} / \partial p^{}_i}$ are tangent to this surface, 
${\bf p} \cdot {\partial {\bf g} \over \partial p^{}_i} = 0$. Hence,    
\begin{equation}
      g^{}_i = - q^{}_{,i} g^{}_3 \, , \quad (i = 1,2) \, ,
  \label{eq028}
\end{equation}
where $q \equiv q(p^{}_1, p^{}_2) \equiv p^{}_3$ denotes the vertical slowness, and 
$q^{}_{,i} \equiv {\partial q} / {\partial p^{}_i}$. Substitution of 
equations~(\ref{eq028}) into equation~(\ref{eq027}) gives a representation 
of the vertical group-velocity component that will be needed later in the 
derivation: 
\begin{equation}
      g^{}_3 = { 1 \over {q - p^{}_1 q^{}_{,1} - p^{}_2 q^{}_{,2}} }.
  \label{eq029}
\end{equation}

Using equations~(\ref{eq028}), we rewrite the horizontal ray displacements 
$x_i$ ($i = 1, 2$) from equations~(\ref{eq026}) as
\begin{equation}
      x_i = - q^{}_{,i} g^{}_3 \tau \, , \quad (i = 1,2) \, .
  \label{eq030}
\end{equation}

Note that $g^{}_3 \tau$ is the depth of the zero-offset reflection point, which 
is independent of the slowness components $(p^{}_1, p^{}_2)$. Therefore, 
differentiating equations~(\ref{eq030}) yields
\begin{equation}
      Y_{i j} \equiv { {\partial x_i} \over {\partial p^{}_j} }  
             = - q^{}_{,ij} g^{}_3 \tau \, ,
  \label{eq031}
\end{equation}
where $q^{}_{,ij} \equiv {\partial^2 q} / {\partial p^{}_i} {\partial p^{}_j}$ is a symmetric matrix of the second derivatives of the vertical slowness.

The NMO ellipse is determined by the matrix ${\bf W}$ 
[equation~(\ref{eq0091})],
\begin{equation}
      {\bf W} = \tau_0 {\bf Y}^{-1} \, , 
 \label{eq132}
\end{equation}
where the inverse matrix ${\bf Y}^{-1}$ should be evaluated for the 
horizontal slowness components of the zero-offset ray.

Substituting $Y_{i j}$ from equation~(\ref{eq031}) into equation~(\ref{eq132}) 
and using expression~(\ref{eq029}) for $g^{}_3$, we obtain 
\begin{equation}
      {\bf W} = \tau_0 {\bf Y}^{-1} =
            { {p^{}_1 q^{}_{,1} + p^{}_2 q^{}_{,2} - q} \over 
              {q^{}_{,11} q^{}_{,22} - q_{,12}^2} } \, 
          \begin{pmatrix}
              q^{}_{,22} & - q^{}_{,12} \cr
          	- q^{}_{,12} &   q^{}_{,11} \cr 
          \end{pmatrix} \! .
  \label{eq032}
\end{equation}

With the matrix ${\bf W}$ from equation~(\ref{eq032}), equation~(\ref{eq009}) 
of the NMO ellipse in a homogeneous arbitrary anisotropic layer takes the 
following form:
\begin{eqnarray}
 V_{\rm nmo}^{-2} (\alpha) & \equiv & V_{\rm nmo}^{-2} (\alpha, p^{}_1, p^{}_2)
  \nonumber \\ & = & {{p^{}_1 q^{}_{,1} + p^{}_2 q^{}_{,2} -q} \over     {q^{}_{,11} q^{}_{,22} - q_{,12}^2} } \, \left[ q^{}_{,22} \cos^2 \alpha - 2 q^{}_{,12} \sin \alpha
    \cos \alpha + q^{}_{,11} \sin^2 \alpha \right] .
  \label{eq633}
\end{eqnarray}

\section{Relation between the exact and rms-averaged NMO velocity} 

Here, we examine the accuracy of the rms averaging of the interval NMO 
velocities for a model that consists of a stack of horizontal arbitrary 
anisotropic layers above a dipping reflector. The interval NMO velocity 
in the $\ell$-th layer is given by equation~(\ref{eq009}): 
\begin{equation}
      V_{{\rm nmo},\ell}^{-2} (\alpha) = 
                 W_{11,\ell} \cos^2 \alpha +
            2 \, W_{12,\ell} \sin \alpha \cos \alpha +
                 W_{22,\ell} \sin^2 \alpha \, .
  \label{d001}
\end{equation}
The symmetric matrix ${\bf W}_\ell$ is expressed through its eigenvalues 
$\lambda_{1,\ell}$ and $\lambda_{2,\ell}$ in equations
(\ref{b002}) -- (\ref{b004}). Here, we assume that 
$\lambda_{1,\ell} > \lambda_{2,\ell}$:
\[
      \lambda_{2,\ell} \equiv \lambda_\ell \, ,
\]
\begin{equation}
      \lambda_{1,\ell} \equiv \lambda_\ell ( 1 + \mu_\ell) \, ,
  \label{d0011}
\end{equation}
where
\[
      \mu_\ell > 0  \, ,
\]
for all $\ell$.

An approximate NMO velocity is obtained by rms averaging of the interval values 
at the azimuth $\alpha$ [equation~(\ref{d001})] as 
\begin{equation}
      \tilde V_{\rm nmo}^2(L,\alpha) = {1 \over {\tau(L)} }
                 \sum_{\ell=1}^L \tau_\ell \, 
                 \left[ W_{11,\ell} \cos^2 \alpha + 
                   2 \, W_{12,\ell} \sin \alpha \cos \alpha +
                        W_{22,\ell} \sin^2 \alpha \right]^{-1} 
  \label{d002}
\end{equation}
In general, $\tilde V_{\rm nmo}(L,\alpha)$ from equation~(\ref{d002}) may be thought of as an approximation of the exact normal-moveout velocity 
$V_{\rm nmo}(L,\alpha)$ from equation~(\ref{eq009}), 
   \begin{eqnarray}
      V_{\rm nmo}^2(L,\alpha) 
        &=& \left[ \,
                 W_{11}(L) \cos^2 \alpha +
            2 \, W_{12}(L) \sin \alpha \cos \alpha +
                 W_{22}(L) \sin^2 \alpha \, \right]^{-1} \nonumber \\
        &=& \left[ \, W_{11}^{-1}(L) W_{22}^{-1}(L) - 
                     (W_{12}^{-1}(L))^2 \, \right] \nonumber \\
        & & \times \left[ \,
                 W_{22}^{-1}(L) \cos^2 \alpha -
            2 \, W_{12}^{-1}(L) \sin \alpha \cos \alpha +
                 W_{11}^{-1}(L) \sin^2 \alpha \, \right]^{-1}  , 
  \label{d003}  
   \end{eqnarray}
where $W_{i j}^{-1}(L)$ are the elements of the inverse matrix 
${\bf W}^{-1}(L)$ given by the Dix-type equation~(\ref{dixavr}):
\begin{equation}
      {\bf W}^{-1}(L) = { 1 \over {\tau(L)} } \, 
             \sum_{\ell=1}^L \tau_\ell \, {\bf W}_\ell^{-1} \, .
  \label{d004}
\end{equation}

Clearly, the direct rms averaging of NMO velocities in equation~(\ref{d002}) 
is different from the more complicated averaging of the inverse matrices 
${\bf W}_\ell^{-1}$ [equation~(\ref{d004})] used to obtain the exact NMO 
velocity in equation~(\ref{d003}). Nevertheless, we will show that the two 
representations of the NMO velocity become identical in the 
{\it linear} approximation with respect to $\mu_\ell$, i.e.,
\begin{equation}
  \tilde V_{\rm nmo}(L,\alpha) = V_{\rm nmo}(L,\alpha) + O(\mu^2_\ell) \, .
  \label{d005}
\end{equation}

In the following derivation, we keep only terms independent of or linear 
in $\mu_\ell$. Combining equations~(\ref{d0011}) 
and~(\ref{b002}) -- (\ref{b004}) allows us to express the interval matrices 
${\bf W_{\ell}}$ through the eigenvalue $\lambda_\ell$ and $\mu_\ell$, 
\[
      W_{11,\ell} = \lambda_\ell \, ( 1 + \mu_\ell \cos^2 \beta_\ell ) \, ,
\]
\begin{equation}
      W_{12,\ell} = \lambda_\ell \, \mu_\ell \sin \beta_\ell 
                                             \cos \beta_\ell \, ,
  \label{d006}
\end{equation}
\[
      W_{22,\ell} = \lambda_\ell \, ( 1 + \mu_\ell \sin^2 \beta_\ell ) \, ,
\]
where $\beta_\ell$ are the rotation angles of the interval NMO ellipses 
introduced in Appendix~A.

Substituting equation~(\ref{d006}) into equation (\ref{d002}), we find the 
following linearized (in $\mu_\ell$) expression for the rms-averaged NMO 
velocity:  
\begin{equation}
      \tilde V_{\rm nmo}^2(L,\alpha) = {1 \over {\tau(L)} }     
      \sum_{\ell=1}^L { {\tau_\ell} \over {\lambda_\ell} } \,
               \left[ 1 - \mu_\ell \cos^2 (\alpha - \beta_\ell) \right] \, .
  \label{d007}
\end{equation}

Next, we need to evaluate the effective NMO ellipse [equation~(\ref{d003})] in the same 
approximation. Using equation~(\ref{d006}) and dropping terms quadratic in 
$\mu_\ell$, we represent the inverse matrices ${\bf W}_\ell^{-1}$ as 
\begin{equation}
      {\bf W}_\ell^{-1} = { 1 \over {\lambda_\ell} } \, 
      \begin{pmatrix}
      1 - \mu_\ell \cos^2 \beta_\ell  &  - \mu_\ell \sin \beta_\ell \cos \beta_\ell \cr
      - \mu_\ell \sin \beta_\ell \cos \beta_\ell  & 1 - \mu_\ell \sin^2 \beta_\ell \cr 
      \end{pmatrix} \! .
  \label{d008}
\end{equation}

After averaging the matrices~(\ref{d008}) in accordance with 
equation~(\ref{d004}) and substituting the result into equation~(\ref{d003}), 
we obtain
\begin{equation}
      V_{\rm nmo}^2(L,\alpha) = {1 \over {\tau(L)} }     
      \sum_{\ell=1}^L { {\tau_\ell} \over {\lambda_\ell} } \,
               \left[ 1 - \mu_\ell \cos^2 (\alpha - \beta_\ell) \right] .
  \label{d009}
\end{equation}
Since equations~(\ref{d007}) and~(\ref{d009}) are identical, the rms-averaged 
velocity $\tilde V_{\rm nmo}$ is indeed equal to the exact NMO velocity in 
the linear approximation with respect to $\mu_\ell$ [equation~(\ref{d005})].
Therefore, $\tilde V_{\rm nmo}$ should represent a good approximation in models with small and {\it moderate} values of $\mu_\ell$, for which terms quadratic in $\mu_\ell$ can be ignored. 

\section{The derivatives of group velocity with respect to the    
         ray parameters $\gamma_1$ and $\gamma_2$}

The derivation in this appendix reproduces the result of Kashtan\cite{Kashtan1982}.
Equation~(\ref{eq0191}) contains the derivative of the group-velocity vector 
for the \mbox{$r$-th} mode ($r = 1,2,3$) in the $\ell$-th layer
[${\partial {\bf g}^{(\ell)}} / {\partial \gamma_n} \equiv
  {\partial {\bf g}^{(\ell,r)}} / {\partial \gamma_n}$] with respect to the 
polar and azimuthal angles $\gamma_1$ and $\gamma_2$ 
of the unit wave-normal vector 
${\bf n} = [\sin \gamma_1 \, \cos \gamma_2,$
$\sin \gamma_1 \, \sin\gamma_2, \, \cos \gamma_1]$. 
Using equations~(\ref{eq008b}) and~(\ref{eq0081}), we find 
the following explicit representation: 
\begin{equation}
      { {\partial g_m^{(\ell,r)}} \over {\partial \gamma_n} } =           
        (a_{imkj} + a_{jmki}) \,
           { {\partial A_i^{(\ell,r)}} \over {\partial \gamma_n} } \,
           p_k^{(\ell,r)} \, A_j^{(\ell,r)} +
        a_{imkj} \, A_i^{(\ell,r)} \,
           { {\partial p_k^{(\ell,r)}} \over {\partial \gamma_n} } \, 
           A_j^{(\ell,r)} \, , 
  \label{eq020}
\end{equation}
where the derivative ${\partial p_k^{(\ell,r)}} / {\partial \gamma_n}$ is
defined by equation (\ref{eq0192}). The derivative of the polarization
vector ${\partial A_i^{(\ell,r)}} / {\partial \gamma_n}$ can be found from
equations~(\ref{eq003}) and~(\ref{eq006}) as

\begin{equation}
      { {\partial A_i^{(\ell,r)}} \over {\partial \gamma_n} } =
        \sum_{s = 1 \atop s \ne r}^3 d_{rs} \, A_i^{(\ell,s)} \, ,
        \quad (n=1,2; ~i=1,2,3) \,,
  \label{eq021}
\end{equation}
where 
\begin{equation}
      d_{rs} = { {V_r^2} \over {V_r^2 - V_s^2} } \,
               (a_{imkj} + a_{ikmj}) \,
               A_i^{(\ell,s)} \, 
               { {\partial p_m^{(\ell,r)}} \over {\partial \gamma_n} } \, 
               p_k^{(\ell,r)} \, A_j^{(\ell,r)} \, ,
               \quad (s,r=1,2,3; ~s \ne r) \,.
  \label{eq022}
\end{equation}


\begin{thebibliography}{99}
	
	\bibitem{GajewskiPsencik1987} D. Gajewski and I. P\v{s}en\v{c}\'{\i}k, Geophysical Journal of the Royal Astronomical Society {\bf 91}, 383 (1987).
	
	\bibitem{TsvankinThomsen1994} I. Tsvankin and L. Thomsen, Geophysics {\bf 59}, no. 8, 1290 (1994).
	
	\bibitem{GrechkaTsvankin1998} V. Grechka and I. Tsvankin, Geophysics {\bf 63}, no. 3, 1079 (1998).
	
	\bibitem{Shah1973} P. M. Shah, Geophysics {\bf 38}, 812 (1973).
	
	\bibitem{HubralKrey1980} P. Hubral and T. Krey, {\it Interval velocities from seismic reflection time measurements} (SEG, 1980).
	
	\bibitem{Thomsen1986} L. Thomsen, Geophysics {\bf 51}, 1954 (1986).
	
	\bibitem{Tsvankin1995} I. Tsvankin, Geophysics {\bf 60}, 268 (1995).
	
	\bibitem{AlkhalifahTsvankin1995} T. Alkhalifah and I. Tsvankin, Geophysics {\bf 60}, no. 5, 1550 (1995).
	
	\bibitem{Tsvankin1997} I. Tsvankin, Geophysics {\bf 62}, no. 4, 1292 (1997).
	
	\bibitem{SayersSEG1995} C. M. Sayers, 65th Annual International Meeting, SEG, Expanded Abstracts pp. 340--343 (1995).
	
	\bibitem{CervenyMolotkovPsencik1977} V. \v{C}erveny, I. A. Molotkov, and I. P\v{s}en\v{c}\'{\i}k, {\it Ray method in seismology} (University of Karlova, 1977).
	
	\bibitem{KendallThomson1989} J. M. Kendall and C. J. Thomson, Geophysical Journal International {\bf 99}, 401 (1989).
	
	\bibitem{Haleetal1992} D. Hale, N. R. Hill, and J. Stefani, Geophysics {\bf 57}, no. 11, 1453 (1992).
	
	\bibitem{Cohen1998} J. K. Cohen, Geophysics {\bf 63}, 275 (1998).
	
	\bibitem{Levin1971} F. K. Levin, Geophysics {\bf 36}, 510 (1971).
	
	\bibitem{MenschRasolofosaon1997} T. Mensch and P. Rasolofosaon, Geophysical Journal International {\bf 128}, 43 (1997).
	
	\bibitem{GajewskiPsencik1996} D. Gajewski and I. P\v{s}en\v{c}\'{\i}k, 66th Annual International Meeting, SEG, Expanded Abstracts pp. 1507--1510 (1996).
	
	\bibitem{Dix1955} C. H. Dix, Geophysics {\bf 20}, 68 (1955).
	
	\bibitem{Alkhalifah1997} T. Alkhalifah, Geophysics {\bf 62}, 662 (1997).
	
	\bibitem{AlDajaniTsvankin1996} A. Al-Dajani and I. Tsvankin, 66th Annual International Meeting, SEG, Expanded Abstracts pp. 1495--1498 (1996).
	
	\bibitem{Cerveny1972} V. \v{C}erveny, Geophysical Journal of the Royal Astronomical Society {\bf 29}, 1 (1972).
	
	\bibitem{Kashtan1982} B. M. Kashtan, Problems of Dynamic Theory of Seismic Wave Propagation {\bf 22}, 14 (1982).
	
	\bibitem{TsvankinHTI1997} I. Tsvankin, Geophysics {\bf 62}, 614 (1997).
	
	\bibitem{Corriganetal1996} D. Corrigan, R. Withers, J. Darnall, and T. Skopinski, 66th Annual International Meeting, SEG, Expanded Abstracts pp. 1834--1837 (1996).
	
	\bibitem{WithersCorrigan1997} R. Withers and D. Corrigan, 59th EAGE Conference and Exhibition, E003 (1997).
	
\end{thebibliography}
\end{document}